\def\apj{ApJ}                                       % ApJ
\def\mnras{MNRAS}                                   % MNRAS
\def\lsim{\lower.5ex\hbox{$\; \buildrel < \over \sim \;$}}
\def\gsim{\lower.5ex\hbox{$\; \buildrel > \over \sim \;$}}
\documentclass{aa}
\usepackage{epsfig}

\begin{document}

 \thesaurus{} % 

%%%%%%%%%%%%%%%%%%%%%%%%%%%%%%%%%%%%%%%%%%%%%%%%%%%%%%%%%%%%%%%%
% The titlepage
%%%%%%%%%%%%%%%%%%%%%%%%%%%%%%%%%%%%%%%%%%%%%%%%%%%%%%%%

\title{Neutron-Capture and 2.22 MeV emission in the atmosphere of the
secondary of an X-ray binary}

\author{P.~Jean \inst{1} \and N.~Guessoum\inst{2}}
\institute{Centre d'Etude Spatiale des Rayonnements, CNRS/UPS, 9 avenue du
colonel Roche, 31028 Toulouse, France \\
email: jean@cesr.fr \and American University of Sharjah, College of Arts \&
Sciences, Physics Unit, Sharjah, UAE \\ email: nguessoum@aus.ac.ae}

   \date{Received 11 May 2001; Accepted 8 August 2001}

   \titlerunning{Neutron-Capture and 2.22 MeV emission}

   \maketitle

%%%%%%%%%%%%%%%%%%%%%%%%%%%%%%%%%%%%%%%%%%%%%%%%%%%%%%%%%%%%%%%%%%%%%%%%%%
% Abstract
%%%%%%%%%%%%%%%%%%%%%%%%%%%%%%%%%%%%%%%%%%%%%%%%%%%%%%%%%%%%%%%%%%%%%%%%%%

\begin{abstract}

We consider the production of 2.22 MeV radiation resulting from the
capture of neutrons in the atmosphere of the secondary in an X-ray binary
system, where the neutrons are produced in the accretion disk around the
compact primary star and radiated in all directions. We have considered
several accretion disk models (ADAF, ADIOS, SLE, Uniform-Temperature) and a
varity of parameters (accretion rate, mass of the compact object, mass,
temperature and composition of the secondary star, distance between the two
objects, etc.). The neutron rates are calculated by a network of nuclear
reactions in the accretion disk, and this is handled by a reaction-rate
formulation taking into account the structure equations given by each
accretion model. The processes undergone by the neutrons in the atmosphere
of the companion star are studied in great detail, including
thermalization, elastic and inelastic scatterings, absorption, escape from
the surface, decay, and capture by protons. The radiative transfer of the
2.22 MeV photons is treated separately, taking into consideration the
composition and density of the star's atmosphere.

The final flux of the 2.22 MeV radiation that can be detected from earth is
calculated taking into account the distance to the source, the direction of
observation with respect to the binary system frame, and the rotation of
the source, as this can lead to an observable periodicity in the flux. We
produce phasograms of the 2.22 MeV intensity as well as spectra of the
line, where rotational Doppler shift effects can lead to changes in the
spectra that are measurable by INTEGRAL's spectrometer (SPI).


\end{abstract}

%%%%%%%%%%%%%%%%%%%%%%%%%%%%%%%%%%%%%%%%%%%%%%%%%%%%%%%%%%%%%%%%%%%%%%%%%%
% 1. Introduction
%%%%%%%%%%%%%%%%%%%%%%%%%%%%%%%%%%%%%%%%%%%%%%%%%%%%%%%%%%%%%%%%%%%%%%%%%%

\section{\label{s1}Introduction}

The capture of a neutron by a proton and the subsequent emission of a
photon at 2.223 MeV has been the object of much interest since the dawn of
gamma-ray astrophysics (Fichtel \& Trombka 1981) because it represents a
potentially important spectroscopic tool in the investigation of
high-energy environments such as solar flares, neutron star surfaces or
accretion disks around compact objects. Several studies have been
conducted both from theory and observation, realizing that this line is
the most promising candidate of all nuclear gamma-ray processes.

Solar flares have had the lion's share of works because of the obvious
possibilities of detailed investigation: Hua \& Lingenfelter (1987);
Murphy et al. (1991); Share \& Murphy (1995); Ramaty et al. (1995);
Ramaty, Mandzhavidze, Kozlovzky (1996); and others. Other sources
have also been investigated, although no detection has been confirmed and
established to date, despite several previous announcements (Jacobson
1982; McConnell et al. 1997).

On the theoretical front, Guessoum \& Dermer (1988) conducted a detailed
theoretical investigation in the context of compact binary sources, taking
Cygnus X-1 as a case-study and pointing out the possibility of emission of
a very narrow 2.22 MeV line from the atmosphere of the companion, while
Aharonian \& Sunyaev (1984) had considered a two-temperature accretion
disk and considered the possibility of emission from the disk itself,
where the line would be very broad. Bildsten and co-workers considered the
possibility of emission at the surface of a neutron star, as a result of
bombardment by accreted protons (Bildsten 1991; Bildsten, Salpeter, and
Wasserman 1993). In the same framework, Bykov et al. (1999) recently
considered the possibility of detecting nuclear gamma-ray lines from
accreting objects with INTEGRAL. Vestrand (1989) investigated the 2.22 MeV
emission resulting from very high energy (E $> 10^{12}$ eV) protons
bombarding companion stars of sources such as Cyg~X-3, Vel~X-1, and Her~X-1,
and predicted significant fluxes ($\approx 10^{-4}$ photons cm$^{-2}$
s$^{-1}$). Finally, Guessoum \& Kazanas (1999), interested in Lithium
production by neutron
bombardment of X-ray binary companions' atmospheres and using the ADAF
accretion disk model, estimated the flux of 2.22 MeV line emission from a
source at about 1 kpc and found it generally very low ($~ 10^{-7}$ photons
cm$^{-2}$
s$^{-1}$).

On the observational front, extensive searches of the line have been
conducted with the data from the Solar Maximum Mission (Harris \& Share
1991) and the COMPTEL instrument of the Compton-GRO mission (McConnell et
al. 1997). The latter showed an excess emission at (l,b) = 300$^o$,-30$^o$
at about 3.7 $\sigma$. No obvious counterparts (X-ray binaries or the like)
could be linked to that general direction, although a catalysmic variable
has been suggested as a likely candidate for that emission. Van Dijk (1996)
has compiled a list of upper-limit 2.22 MeV binary-source radiation fluxes
based on COMPTEL observations of 27 black-hole candidates. These upper
limits range from 1 to 5 $\times 10^{-5}$ photons cm$^{-2}$ s$^{-1}$.

The production of neutrons in the accretion disks around the primaries
of X-ray binaries is a direct result of the high temperatures that prevail in
such environments. The gravitational attraction, combined with the viscous
dissipation, leads to a significant heating of the plasma, particularly
its nuclei, resulting in temperatures as high as 10$^{10}$ -- 10$^{12}$ K
in the inner regions (r $\approx 1 - 100$ R$_S$, where R$_S$ = 2GM/c$^2$ is the
Schwarzchild radius of the central compact object). Simple considerations
show that densities of the accretion plasma can range between 10$^{12}$
and 10$^{18}$ cm$^3$ depending on the geometry and the model, which
implies substantial nuclear reactions. Since the accreted material is
expected to contained He and/or metals (depending on whether the secondary
star's composition is normal or of Wolf-Rayet type), breakup reactions
will produce significant amounts of neutrons. Part of the present paper is
devoted to computing precisely such amounts under various assumptions:
different accretion disk models, different initial compositions of the
accreted material, different temperature profiles, etc. This will be
presented in detail in Section 2.

There are several reasons underlying our interest in 2.22 MeV emission
from the secondary stars of X-ray binaries. First and foremost, as
mentioned above the CGRO 2.22 MeV map showed that there are real
possibilities for the detection of this line, which would consitute the
first observation of any nuclear gamma-ray line outside of the solar
system, and this prospect is heightened by the upcoming INTEGRAL mission's
order-of-magnitude improvement in both flux sensitivity and energy
resolution, which become crucial when most predictions of line fluxes from
such sources are at the 10$^{-5}$ photons cm$^{-2}$ s$^{-1}$ level, that is
below the
sensitivity of CGRO and right at the detection threshold of INTEGRAL.
Secondly, the accretion disk problem remains fully unresolved (very little
progress has been made on the determination of the disk structure, much
less the identification of the viscosity mechanism), and since our results
will show significant differences in fluxes and therefore in the possibility
for detection of this line emission, this may constitute further testing
of models. Furthermore, we will find that line fluxes depend
substantially on the secondary star's characteristics (elemental
composition in particular), the eventual detection of such a signal would
then also constitute a diagnostic of the X-ray binary's secondary (e.g.
convection and mixing of the gas).

This paper is structured as follows: in Sect. 2, the treatment of
neutron production in the accretion disk (and its subsequent escape) is
presented for various models; in Sect. 3, the detailed treatment of
neutron propagation, slow-down, and capture in the atmosphere of the
secondary is shown; in Sect. 4 we present the results of line fluxes
when everything is combined: neutron production and escape from the disk,
various interactions and
effects (geometrical as well as physical) in the atmosphere of the
secondary, and finally capture and gamma-ray emission, with star rotation,
photon energy shift, etc.; in the final
section we discuss our results with INTEGRAL observation prospects in mind
and point to future work on the subject.

%%%%%%%%%%%%%%%%%%%%%%%%%%%%%%%%%%%%%%%%%%%%%%%%%%%%%%%%%%%%%%%%%%%%%%%%%%
% 2.  Neutron Production in the Accretion Disk
%%%%%%%%%%%%%%%%%%%%%%%%%%%%%%%%%%%%%%%%%%%%%%%%%%%%%%%%%%%%%%%%%%%%%%%%%%

\section{\label{s2} Neutron Production in the Accretion Disk }

Neutrons are produced by means of nuclear breakup reactions involving
Hydrogen and Helium nuclei. Once produced, the
neutrons can escape or carry part of the angular momentum outward by
collisions with the infalling nuclei, thereby participating in the viscous
dissipation, as they are not affected by the presence of any magnetic
fields. In fact, the escape of the neutrons from the gravitational
attraction of the compact object can be computed quite accurately in the
assumption of thermal conditions (Aharonian \& Sunyaev 1984; Guessoum \&
Kazanas 1990). So, at least in principle, the calculation of the flux of
neutrons irradiating the surface of the companion star is quite feasible.
In practice, however, there are a number of factors which intervene and
make the task somewhat more complicated: the temperature and density
distribution of matter in the accretion disk is model-dependent, the
composition and nuclear abundances of the plasma depends on stellar
evolution conditions, and cross sections for the various reactions are not
always known accurately (and this depends also on the extent of the
nuclear reactions network taken into account in the computation).

Since the nuclear reactions relevant to our problem and to the production
of gamma rays require energies of about 10 MeV per nucleus, our interest
is limited to the inner regions of the accretion flow, i.e.~regions around the
compact object between about 100 Schwarzschild radii) and either the
surface of the neutron star or the horizon of the black hole, depending on
the nature of the compact object. For general purposes, we assume a mass
of $1 \, M_{\odot}$ for the compact object, unless a specific case is
considered.

In this work we consider a number of relatively simple accretion disk
models:

\begin{itemize}

\item{1 -} The Advection-Dominated Accretion Flow (ADAF) model proposed by
Narayan and co-workers (Narayan \& Yi 1994, Chen et al. 1995, Narayan, Yi,
and Mahadevan 1995, Narayan \& Yi 1995), which has the added appeal that its
temperature and density function are very simple and analytical expressions
(Yi \& Narayan 1997):

\begin{equation}
T_i = 1.11 \times 10^{12} \, r^{-1} {\rm K} = 95.68 \, r^{-1} \; {\rm MeV}
\label{eq:n1}
\end{equation}

\noindent where $r$ is the dimensionless radial variable $r = R / R_S$;

\begin{equation}
n = 6.4 \times 10^{18} \; \alpha^{-1} \; m^{-1} \; \dot m \; r^{-3/2} \; {\rm
cm}^{-3}
\label{eq:n2}
\end{equation}

\noindent where $m$ is the dimensionless mass measured in units of a solar
mass ($m = M / M_{\odot}$), $\dot m$ is the dimensionless accretion rate
measured in units of the Eddington accretion rate $\dot m = \dot M / \dot
M_{Edd} = {\dot M} / {1.4 \times 10^{17} m}~~ {\rm g/s}$ (in this work we
will consider values of $\dot M$ in the range of $10^{-10} \ M_{\odot}/yr$
-- $10^{-8} \ M_{\odot}/yr$), taking $\beta$ (the ratio of gas pressure to
total pressure) to equal 1/2 (as in Yi and Narayan 1997); the central
factor $\alpha$ is taken to equal either 0.3 or 0.1 (following Narayan \&
Yi 1994).

Furthermore, it must be noted that in this model the value of the electron
temperature does not affect our results, which depend mainly on the profile
of the ion temperature T$_i$(r). 

\medskip

\item{2 -} The Advection-Dominated Inflow-Outflow Solutions proposed by
Blandford and Begelman (1998), which consists of a ``generalized" ADAF
model where only a small fraction of the gas initially supplied ends up
falling onto the compact object, and where the energy released is
transported radially outward and effectively becomes a wind, driving away
the rest of the accreted material. 

In this family of solutions, a central assumption is the radial dependence of the accretion rate as ${\dot m(r) \propto r^p, \, 0 \le p < 1 \; }$.  In the present work, we take the ``gasdynamical wind" (case iv of Blandford \& Begelman 1998) solution, i.e. $p = 1/2, \, \epsilon = 1/2$, which leads to a density function $ n \propto r^{-1}$; more precisely:

\begin{equation}
n = 2.18 \times 10^{19} \; \alpha^{-1} \; \frac{(4.52 \times 10^{8}) \times \dot M}
{m r} \; cm^{-3} \; .
\label{eq:n3}
\end{equation}

Note that this special form of the density function also corresponds to the solution obtained by Kazanas, Hua, \& Titarchuk (1997) in fitting the time lags in similar, compact X-ray accreting sources. 

Another important equation is that of the disk height h (in units of R$_S$), which in the particular solution adopted here, is given by:

\begin{equation}
h = 0.447 \, \times 3 \, m \, r  \; ;
\label{eq:n4}
\end{equation}

Also, the radial velocity is

\begin{equation}
v_r = 0.52 \; \alpha \; r^{- \frac{1}{2}} \; c \; cm/s \; .
\label{eq:n5}
\end{equation}

Finally, the ion temperature profile is the same as in ADAF.

\medskip

\item{3 -} The Two-Temperature Shapiro-Lightman-Eardley (1976) Model, for
which the expressions are given by:

\begin{equation}
T_i = 5 \times 10^{11} \; M_*^{- \frac{5}{6}} \, \dot M_*^{\frac{5}{6}} \,
\alpha^{- \frac{7}{6}} \, \phi^{\frac{5}{6}} \, r_*^{- \frac{5}{4}} \; K
\label{eq:n6}
\end{equation}

\begin{equation}
T_e = 7 \times 10^8 \; M_*^{\frac{1}{6}} \, \dot M_*^{- \frac{1}{6}} \,
\alpha^{- \frac{1}{6}} \, \phi^{- \frac{1}{6}} \, r_*^{\frac{1}{4}} \; K
\label{eq:n7}
\end{equation}

\begin{equation}
h = 10^5 \; M_*^{\frac{7}{12}} \, \dot M_*^{- \frac{5}{12}}
\, \alpha^{- \frac{7}{12}} \, \phi^{\frac{7}{12}} \, r_*^{\frac{7}{8}} \; cm
\label{eq:n8}
\end{equation}

\begin{equation}
\rho = 5 \times 10^{-5} \; M_*^{- \frac{3}{4}} \, \dot M_*^{-
\frac{1}{4}} \, \alpha^{\frac{3}{4}} \, \phi^{- \frac{1}{4}} \, r_*^{-
\frac{9}{8}} \; g/cm^3 \; ,
\label{eq:n9}
\end{equation}

\noindent where $M_* = M /3 ; \phi = 1 -\sqrt{3R_S/R} = 1 - \sqrt{3/r} ;
\dot M_* = \dot M / 10^{17} g/s ; r_* = 2R/R_S = 2 r \; .$

\medskip

\item{4 -} The uniform-temperature model, similar to the one used in the
Guessoum \&
Kazanas (1990) neutron-viscosity model, where T$_i$ is set by a viscosity
condition but is taken to be uniform in the disk, and T$_e$ is chosen
arbitrarily (usually between 0.1 and 0.5 MeV); the density is then
obtained from an energy balance requirement (viscous heating of the ions
equal to Coulomb energy transfer from the ions to the electrons). In the
Guessoum \& Kazanas model, the following relations were derived:

\begin{equation}
n = \frac{3.375 \times 10^{25}}{r^2} \sqrt{\frac{\dot m}{m^3 \, f_{th}}}
\label{eq:n10}
\end{equation}

\noindent where $f_{th}$ is the ion-electron Coulomb energy transfer
function (Dermer 1986),

\noindent and

\begin{equation}
t_{visc} = \frac{4\pi}{3} m_i \, R^3 \frac{n}{\dot m}
\label{eq:n11}
\end{equation}

\noindent where $m_i$ is the mass of the ion.

\end{itemize}

\medskip

Finally we note that in each case the system is taken to be in
steady-state, whereby the accretion and in-fall are constant in time, and
thus the neutron production and gamma-ray emission are steady.

%%%%%%%%%%%%%%%%%%%%%%%%%%%%%%%%%%%%%%%%%%%%%%%%%%%%%%%%%%%%%%%%%%%%%%%%%%
\subsection{\label{s21} The Nuclear Reactions}

Assuming an initial composition for the accreting plasma, that is some
initial fractions of Hydrogen and Helium (90\% H, 10\% He in the ``normal"
case, and 10\% H, 90\% He in the ``helium-rich case", which would
correspond to a Wolf-Rayet secondary star companion), we compute the rates
of the main nuclear reactions which the ions (the protons, alphas, $^3$He
and $^2$H nuclei) can undergo:
$$p + \alpha  \longrightarrow {}^3He + {}^2H $$
$$\alpha + \alpha \longrightarrow {}^7Li + p $$
$$p + {}^3He  \longrightarrow 3p + n $$
$$\alpha + \alpha \longrightarrow {}^7Be + n $$
$$\alpha + {}^3He  \longrightarrow \alpha + 2p + n $$
$$p + {}^2H  \longrightarrow 2p + n $$

The first two reactions contributing to the destruction of Helium (our
source of neutrons), and the last four reactions are the actual
neutron-producing processes. We disregard all high-energy
neutron-production from proton-proton collisions. The cross sections are
essentially the same as those used in Guessoum \& Kazanas (1999), and the
reaction rates are calculated through the usual non-relativistic
expression for binary processes ($r_{ij} = n_i n_j/(1 + \delta_{ij})
<\sigma_{ij} v_{ij}>$, the averaging here is performed over thermal
distributions for the ions).

We then let the plasma evolve over the dynamical timescale $t_d$, with its
temperature and density set by the disk structure equations given above
for each model, as the material sinks in the gravitational well.  We use a
simple numerical scheme of explicit finite-differencing to follow the
abundances of the various species (the neutrons in particular). This
calculation is performed for various values of the model parameters: $\dot
M = 10^{-10}, \; 10^{-9} \;$ and $10^{-8} \; M_{\odot}/ {\rm yr}$; $\alpha
= 0.3 $ and $ 0.1$; $M = 1 $; etc. Note that since we are only interested
in neutron production in the disk and not in any radiative processes, the
electron temperature T$_e$, whenever needed, is taken as a free parameter,
usually equal to either 100 or 500 keV.

The escape of the neutrons is handled in the same way as in Aharonian \&
Sunyaev (1984) or Guessoum \& Kazanas (1990), that is by computing the
fraction of neutrons that have kinetic energies sufficient to overcome the
gravitational binding to the compact object:  $ 1/2 m_n (\vec{v_{flow}} +
\vec{v_{thermal}})^2 > G M m_n / R $; this translates into a fraction of
escaping neutrons given by Eqs. (13) -- (15) of Guessoum \& Kazanas
(1990).

One final important aspect of the neutrons produced is their energy
distribution. This aspect was addressed briefly in Guessoum \& Kazanas
(1999), where it was concluded that a thermal distribution of the neutrons
with a temperature equal to that of the ions in the disk is reasonable.
This point proves to be important not only for the escape fraction, but
also for the interaction of those neutrons that reach the secondary star's
atmosphere, since their various interactions there depend strongly on
their kinetic energies.

%%%%%%%%%%%%%%%%%%%%%%%%%%%%%%%%%%%%%%%%%%%%%%%%%%%%%%%%%%%%%%%%%%%%%%%%%%
\subsection{\label{s22} Neutron Production Results in Various Disks}

\begin{figure*}
    \epsfig{file=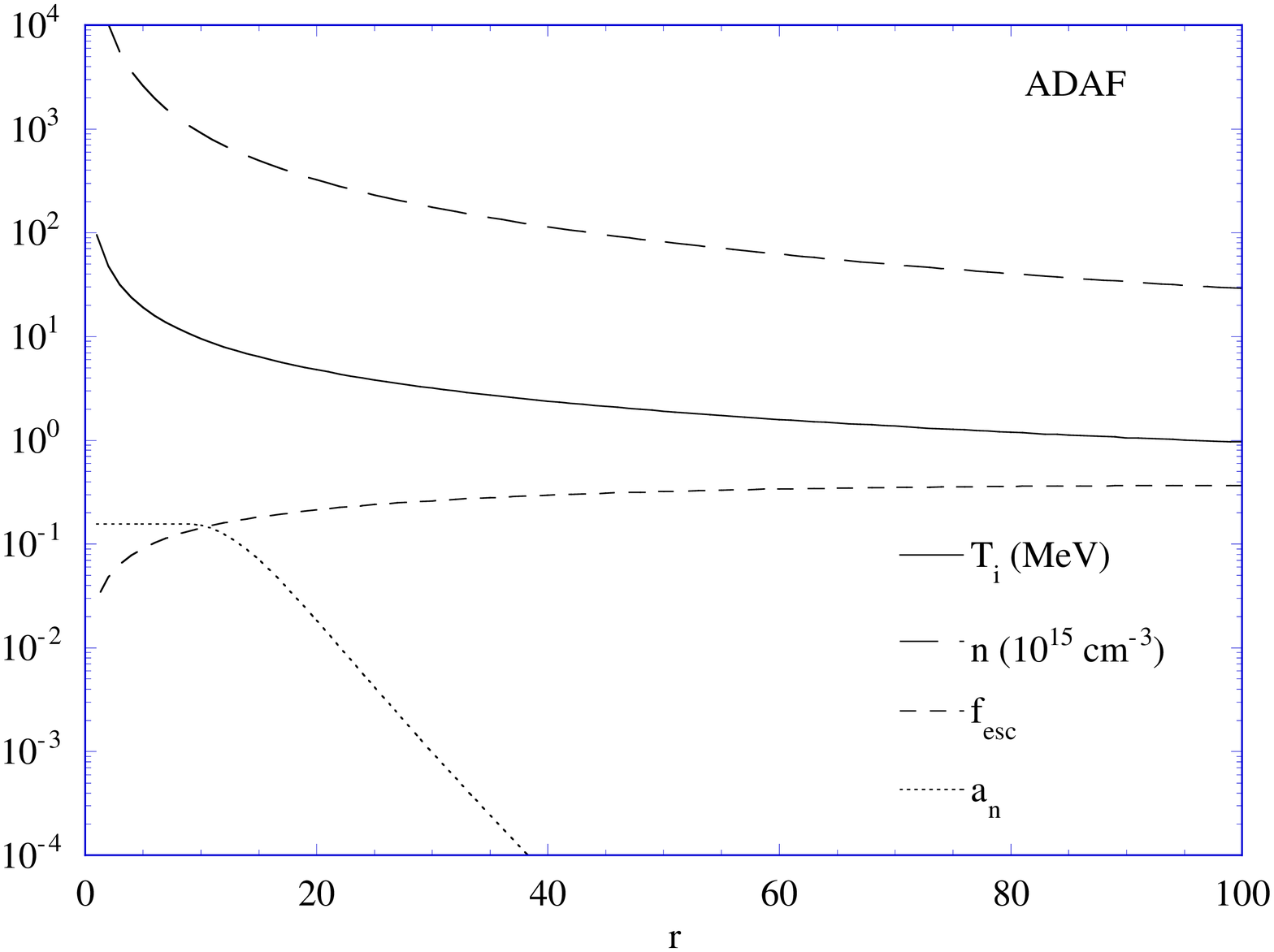,height=6.1cm,width=8.6cm}
    \epsfig{file=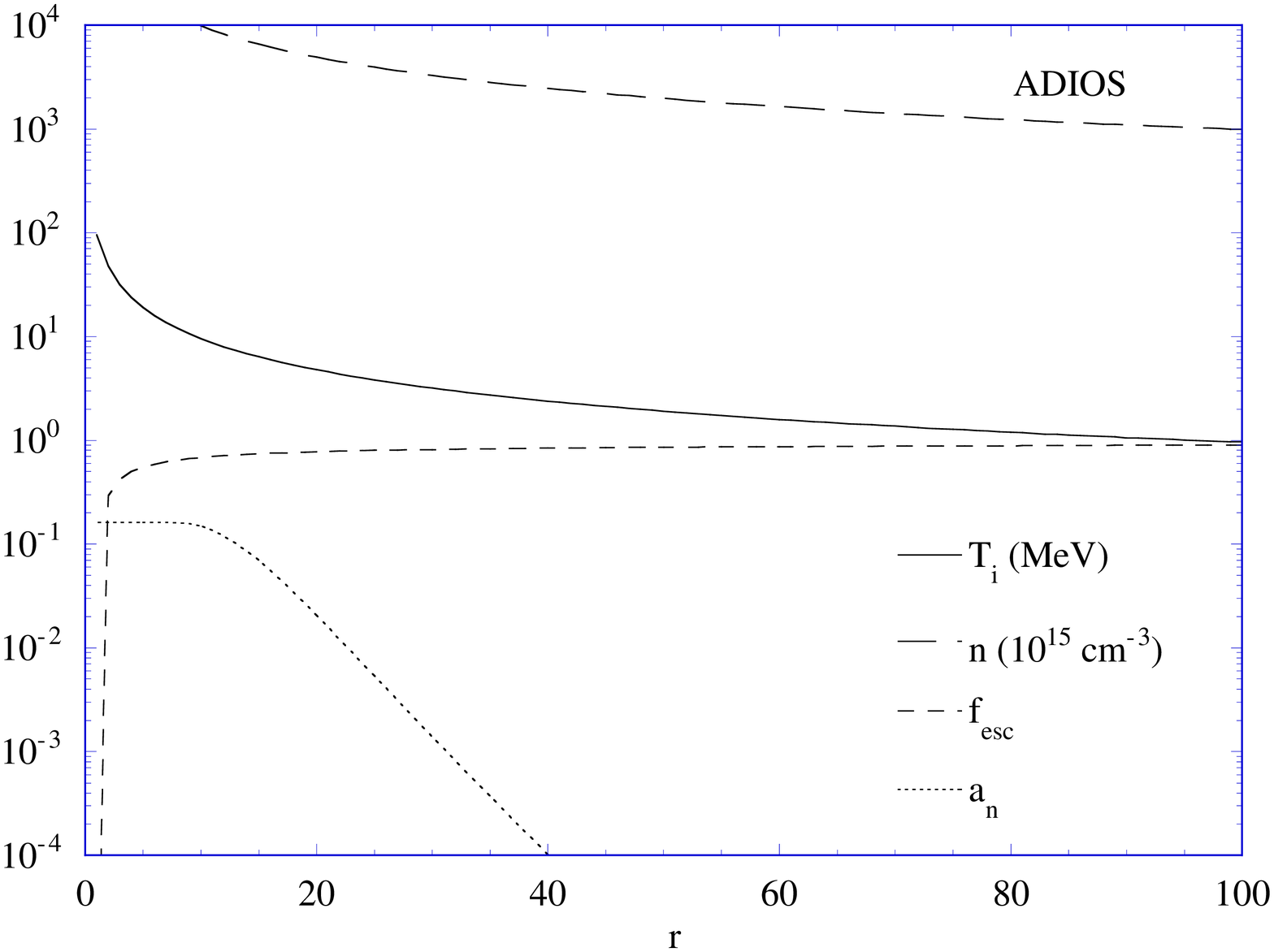,height=6.1cm,width=8.6cm}
    \epsfig{file=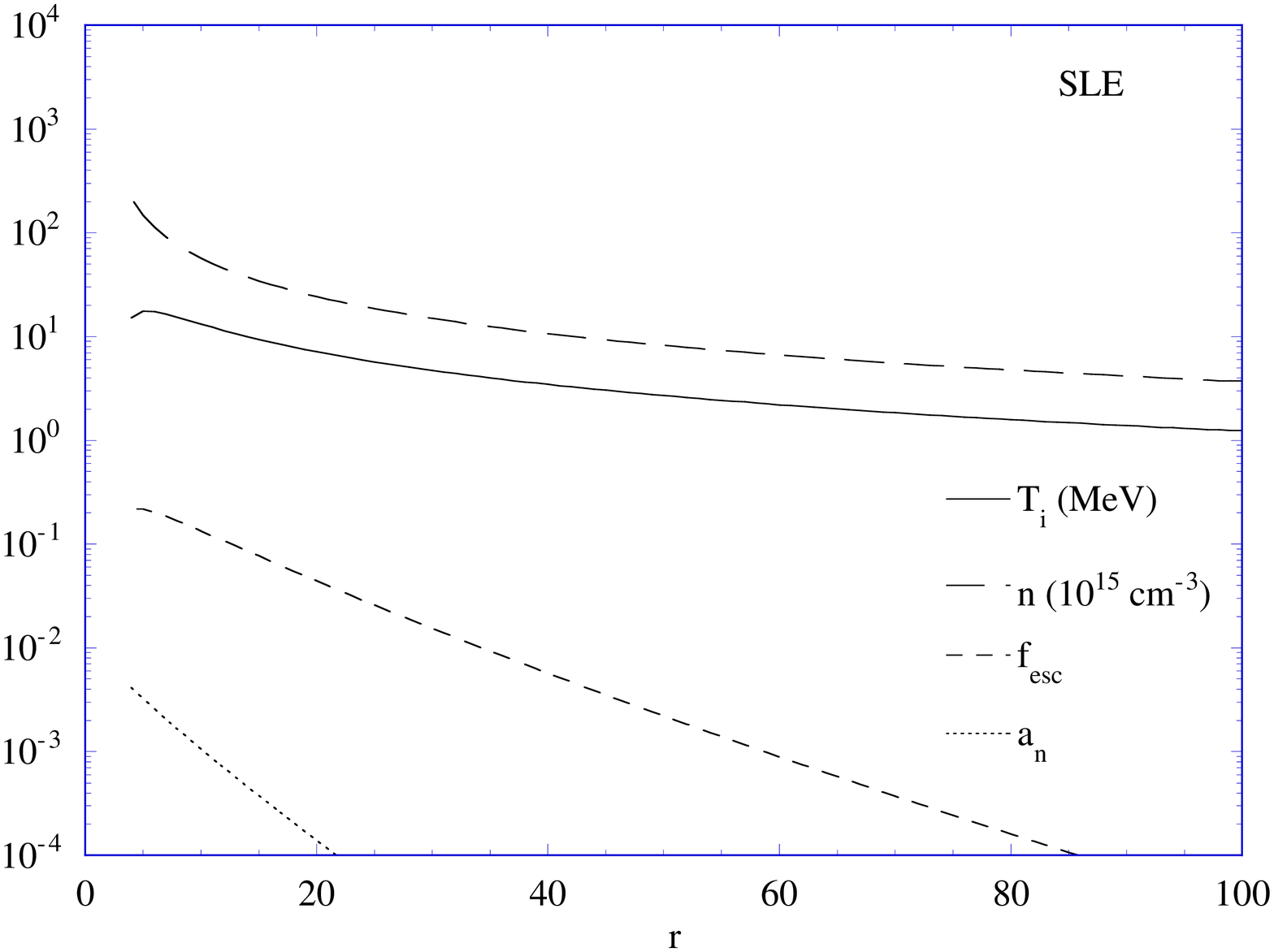,height=6.1cm,width=8.6cm}
    \epsfig{file=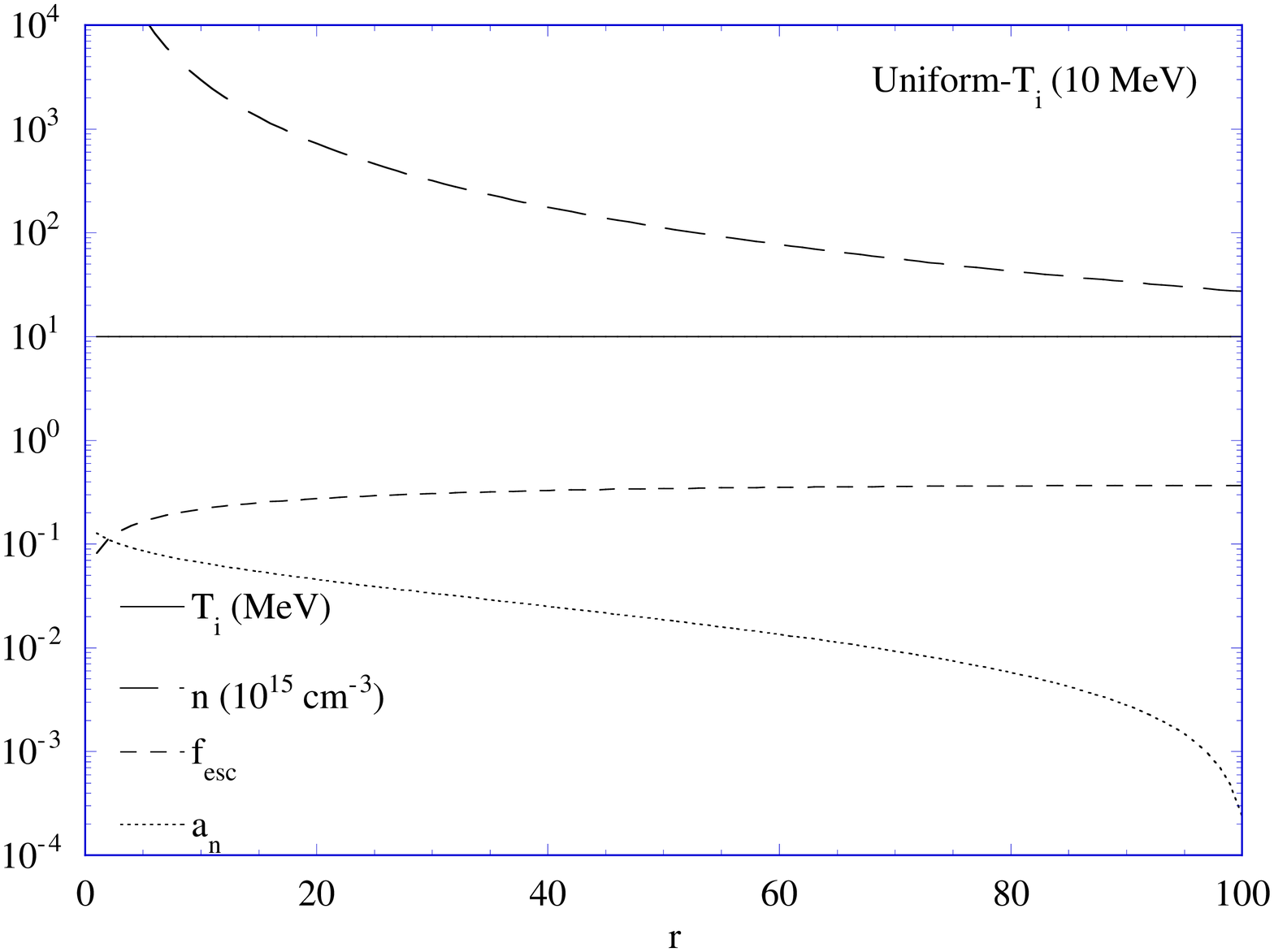,height=6.1cm,width=8.6cm}
    \epsfig{file=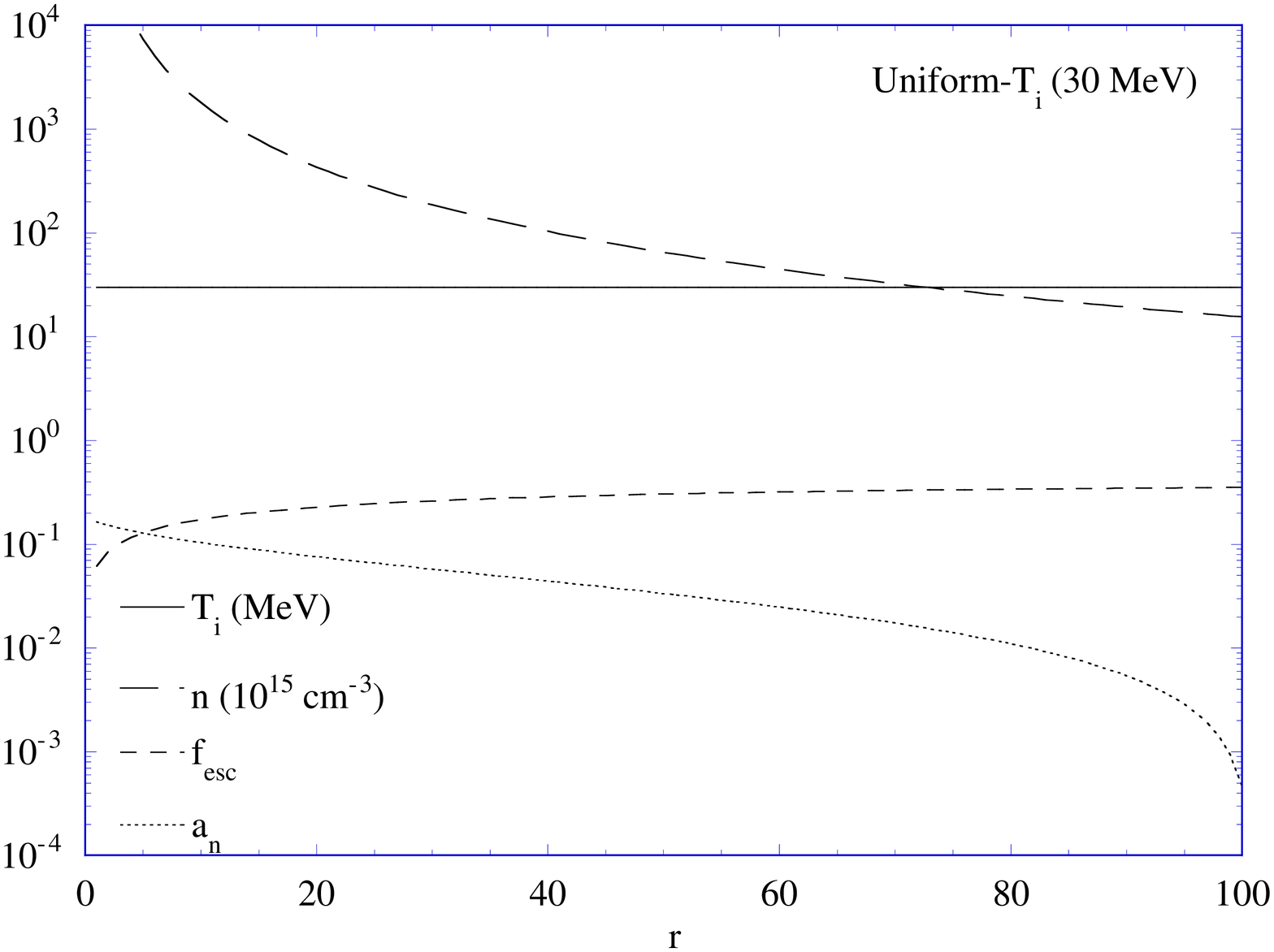,height=6.1cm,width=8.6cm}
   \caption{Plasma characteristics (density, ion temperature, escape
fraction of neutrons, and abundance of neutrons) as a function of r, in the
ADAF, ADIOS, SLE, and Uniform-T$_i$ (10 and 30 MeV), assuming ${\dot M} = 10^{-9} M_{\odot}$ /yr, $\alpha = 0.1$, and $T_e = 0.1$ MeV.}
       \label{fig:plasma}
   \end{figure*}

Before presenting results of neutron production for the various disk
models, we first show the main characteristics of these models. Figures
\ref{fig:plasma} show the plasma density n (in units of 10$^{15}$ cm$^{-3}$),
the ion temperature T$_i$ (in units of MeV), the escape fraction of
neutrons f$_{esc}$, and the abundance of neutrons achieved in the disk,
all as a function of r = R / R$_S$, in the ADAF, ADIOS, SLE, and Uniform-T$_i$
(10 and 30 MeV), respectively, for disks with solar composition (90\% H and
10\% He)
and an accretion rate of $10^{-9}$ M$_{\odot}$/yr; for ADAF, ADIOS, and SLE
the
value of $\alpha$ in the figures is 0.1, and for the uniform-T$_i$ models the
electron-temperature used for the figures is 0.5 MeV.

%%%%%%%% Figure: Plasma characteristics %%%%%%%%
%%%%%%%%%%%%%%%%%%%%%%%%%%%%%%%%%%%%%%%%%%%%%%%%%%%%%%%%%%%%%%%%%%%%%%%

Table 1 presents the results of these calculations, showing the
rates of neutron production in the ADAF-model case for various values of
the viscosity parameter $\alpha$ and the mass accretion rate $\dot M$ (1a
corresponds to a ``normal" composition; 1b corresponds to the ``Helium-rich"
case). All the results shown correspond to $ M = 1 \, M_{\odot}$ and $R
\approx 100 \ R_S \approx 10^7$ cm, though the actual results are
mostly insensitive to the value of the outer radius of the disk, as most
of the nuclear reactions and neutron production take place in the inner
part (less than about 10 $R_S$) of the accretion flow where the ion temperature
is several tens of MeV.

Table 2 is equivalent to Table 1 in the ADIOS model.
Table 3 corresponds to the SLE model; Tables 4 and 5 correspond
to the uniform-T$_i$ model, with T$_i$ = 30 MeV and T$_i$ = 10 MeV.
Respectively.

It is interesting to note that the resulting neutron fluxes depend
strongly on the models considered, but also are quite sensitive to the
3 disk parameters considered, namely $\alpha$, $\dot M$, and the initial
composition of the plasma.

\begin{table}
    \caption{Neutron Production Rate (in neutrons/s) in the ADAF Model, for
initial (a) 90\% H and 10\% He and (b) 10\% H and 90\% He compositions.}
    \begin{array}[b]{lcc}
 \hline
 \noalign{\smallskip}
 \dot M \, (M_{\odot}/yr) & \; \; \; \; \; \; \; \; \; \; \; \; \; \; \; \;
\; a \; \; \; \; \; \; \; \; \; \; \; \; \; \; \; \; \; \; \; \;  & \; \;
\; \; \; \; \; \; \; \; \; \; \; b  \; \; \; \; \; \; \; \; \; \; \; \; \;
\; \; \\
   \end{array}
    \begin{array}[b]{lcccc}
%\noalign{\smallskip}
\; \; \; \; \; \; \; \; \; \; \; \; \; \; \; \; \; \; \; \; & \alpha = 0.3
& \alpha = 0.1 & \alpha = 0.3 & \alpha = 0.1 \\
 \hline
 \noalign{\smallskip}
10^{-10} &  1.1\times 10^{37} & 5.6\times 10^{37} & 6.6\times 10^{37} &
2.7\times 10^{38} \\
10^{-9}  &  4.4\times 10^{38} & 1.4\times 10^{39} & 2.6\times 10^{39} &
6.0\times 10^{39} \\
10^{-8} &  9.1\times 10^{39} & 1.6\times 10^{40} & 3.9\times 10^{40} &
7.3\times 10^{40} \\
\noalign{\smallskip}
\hline
    \end{array}
  \label{tab:1}
\end{table}

\begin{table}
    \caption{Neutron Production Rate in the ADIOS Model, for initial (a)
90\% H and 10\% He and (b) 10\% H and 90\% He compositions.}
    \begin{array}[b]{lcc}
 \hline
 \noalign{\smallskip}
 \dot M \, (M_{\odot}/yr) & \; \; \; \; \; \; \; \; \; \; \; \; \; \; \; \;
\; a \; \; \; \; \; \; \; \; \; \; \; \; \;\; \; \; \; \; \; \;  & \; \; \;
\; \; \; \; \; \; \; \; \; \; b  \; \; \; \; \; \; \; \; \; \; \; \; \; \;
\; \\
   \end{array}
    \begin{array}[b]{lcccc}
%\noalign{\smallskip}
\; \; \; \; \; \; \; \; \; \; \; \; \; \; \; \; \; \; \; \; & \alpha = 0.3
& \alpha = 0.1 & \alpha = 0.3 & \alpha = 0.1 \\
 \hline
 \noalign{\smallskip}
10^{-10} & 1.7\times 10^{37} & 1.4\times 10^{38} &  1.4\times 10^{38} &
7.9\times 10^{38} \\
10^{-9} & 1.6\times 10^{39}  & 7.8\times 10^{39} &  9.9\times 10^{39}  &
3.5\times 10^{40} \\
10^{-8}  & 8.3\times 10^{40} & 2.3\times 10^{41} &  3.6\times 10^{41} &
9.8\times 10^{41} \\
\noalign{\smallskip}
\hline
    \end{array}
  \label{tab:2}
\end{table}

\begin{table}
    \caption{Neutron Production Rate in the SLE Model, for initial (a) 90\%
H and 10\% He and (b) 10\% H and 90\% He compositions.}
    \begin{array}[b]{lcc}
 \hline
 \noalign{\smallskip}
 \dot M \, (M_{\odot}/yr) & \; \; \; \; \; \; \; \; \; \; \; \; \; \; \; \;
\; a \; \; \; \; \; \; \; \; \; \; \; \; \; \; \; \; \; \; \; \;  & \; \;
\; \; \; \; \; \; \; \; \; \; \; b  \; \; \; \; \; \; \; \; \; \; \; \; \;
\; \; \\
   \end{array}
    \begin{array}[b]{lcccc}
%\noalign{\smallskip}
\; \; \; \; \; \; \; \; \; \; \; \; \; \; \; \; \; \; \; \; & \alpha = 0.3
& \alpha = 0.1 & \alpha = 0.3 & \alpha = 0.1 \\
 \hline
 \noalign{\smallskip}
10^{-10} & 3.6\times 10^{23} & 5.0\times 10^{33} &  6.5\times 10^{24} &
8.6\times 10^{34}  \\
10^{-9} & 4.2\times 10^{36}  & 2.9\times 10^{38} &  5.0\times 10^{37}  &
1.8\times 10^{39} \\
10^{-8} & 5.0\times 10^{39} & 1.9\times 10^{40} &  3.3\times 10^{40} &
1.3\times 10^{41} \\
\noalign{\smallskip}
\hline
    \end{array}
  \label{tab:3}
\end{table}

\begin{table}
    \caption{Neutron Production Rate in the Uniform Ion Temperature Model
with kT$_i$ = 30 MeV , for initial (a) 90\% H and 10\% He and (b) 10\% H
and 90\% He compositions.}
    \begin{array}[b]{lcc}
 \hline
 \noalign{\smallskip}
 \dot M \, (M_{\odot}/yr) & \; \; \; \; \; \; \; \; \; \; \; \; \; \; \; \;
\; a \; \; \; \; \; \; \; \; \; \; \; \; \; \; \; \; \; \; \; \;  & \; \;
\; \; \; \; \; \; \; \; \; \; \; b  \; \; \; \; \; \; \; \; \; \; \; \; \;
\; \; \\
   \end{array}
    \begin{array}[b]{lcccc}
%\noalign{\smallskip}
\; \; \; \; \; \; \; \; \; \; \; \; \; \; \; \; \; \; \; \; &  kT_e= &
kT_e= &  kT_e= & kT_e=  \\
\; \; \; \; \; \; \; \; \; \; \; \; \; \; \; \; \; \; \; \; & 0.5 MeV & 0.1
MeV & 0.5 MeV & 0.1 MeV \\
\hline
 \noalign{\smallskip}
10^{-10} &  3.2\times 10^{38}  &   4.8\times 10^{37} &  7.4\times 10^{38}
&   1.0\times 10^{38} \\
10^{-9} &  2.4\times 10^{39}  &   4.3\times 10^{38} &  6.0\times 10^{39}  &
1.0\times 10^{39} \\
10^{-8} &  1.6\times 10^{40}  &   5.1\times 10^{39} &  4.4\times 10^{40}  &
3.1\times 10^{40} \\
\noalign{\smallskip}
\hline
    \end{array}
  \label{tab:4}
\end{table}

\begin{table}
    \caption{Neutron Production Rate in the Uniform Ion Temperature Model
with kT$_i$ = 10 MeV , for initial (a) 90\% H and 10\% He and (b) 10\% H
and 90\% He compositions.}
    \begin{array}[b]{lcc}
 \hline
 \noalign{\smallskip}
 \dot M \, (M_{\odot}/yr) & \; \; \; \; \; \; \; \; \; \; \; \; \; \; \; \;
\; a \; \; \; \; \; \; \; \; \; \; \; \; \; \; \; \; \; \; \; \;  & \; \;
\; \; \; \; \; \; \; \; \; \; \; b  \; \; \; \; \; \; \; \; \; \; \; \; \;
\; \; \\
   \end{array}
    \begin{array}[b]{lcccc}
%\noalign{\smallskip}
\; \; \; \; \; \; \; \; \; \; \; \; \; \; \; \; \; \; \; \; &  kT_e=  &
kT_e= &  kT_e=  & kT_e= \\
 \; \; \; \; \; \; \; \; \; \; \; \; \; \; \; \; \; \; \; \; &  0.5 MeV  &
0.1 MeV &  0.5 MeV  & 0.1 MeV \\
\hline
 \noalign{\smallskip}
10^{-10} &  2.0\times 10^{38}  &   2.1\times 10^{37} &  4.0\times 10^{38}
&   4.1\times 10^{37}\\
10^{-9} &  1.6\times 10^{39}  &   2.1\times 10^{38} &  3.6\times 10^{39}  &
4.8\times 10^{38} \\
10^{-8} &  1.1\times 10^{40}  &   1.8\times 10^{39} &  2.7\times 10^{40}  &
8.0\times 10^{39} \\
\noalign{\smallskip}
\hline
    \end{array}
  \label{tab:5}
\end{table}

%%%%%%%%%%%%%%%%%%%%%%%%%%%%%%%%%%%%%%%%%%%%%%%%%%%%%%%%%%%%%%%%%%%%%%%%%%
% 3. Propagation, slow-down and capture of neutrons
%%%%%%%%%%%%%%%%%%%%%%%%%%%%%%%%%%%%%%%%%%%%%%%%%%%%%%%%%%%%%%%%%%%%%%%%%%
\section{\label{s3} Propagation, slow-down and capture of neutrons in the
atmosphere of the secondary}

A fraction of the neutron flux emitted by the accretion disk irradiates
the atmosphere of the secondary star. Some of these neutrons
get thermalized; some are captured by nuclei, others decay or
escape the secondary if after several scatterings their kinetic energies
become larger than the gravitational potential energy binding them to
the star. The 2.22 MeV photons resulting from the capture of neutrons by
protons escape the secondary atmosphere if they are not absorbed or
Compton scattered in the surrounding gas. Consequently, the probability of
escape of 2.22 MeV photons depends on the depth of their creation site (in
the atmosphere) and on their direction of emission with respect to the
surface.

These processes have all been modeled, in the aim of determining the rate
of 2.22 MeV radiation emitted by the secondary. The transport of neutrons
in the secondary has been simulated using the code GEANT/GCALOR, which
takes into account elastic and inelastic interactions as well as the
neutrons' decay and capture by nuclei. The radiative transfer of 2.22 MeV
photons in the atmosphere and the fraction of neutrons falling back onto
the secondary are then modeled separately, using the results of the
previous simulations. The total emissivity at 2.22 MeV is estimated by
integrating over the whole secondary surface irradiated by neutrons. In
the following paragraphs details of the method are presented.

The secondary star has a mass and a radius denoted hereafter by M$_s$ and
R$_s$. (There should be no confusion between the Schwarzchild radius (R$_S$) of the
primary and the radius (R$_s$) of the secondary star.) The star's atmosphere is
characterized by a temperature T$_s$, a density $\rho_s$ and
an abundance in H and in He of [H] and [He], respectively. The isotopic
abundance of these elements are taken from Anders and Grevesse (1989). We
have disregarded all other elements in the atmosphere. The distance
between the accretion disk and the center of the secondary (also called
separation) is parametrized by d$_s$. This distance d$_s$ and the radius of
the star determine the fraction of the neutron flux emitted by the
accretion disk that irradiates the secondary. The secondary's mass and
radius allow us to calculate the gravitational potential and therefore
the fraction of escaping neutrons, which also depends on their energy and
direction. The temperature, the density, and the composition of the
atmosphere are parameters for the simulation of neutron tracks in the
secondary. The abundance of $^3$He is a crucial parameter for the
intensity of the 2.22 MeV line emission since, though it is estimated to
be $\approx 10^5$ less abundant than H, it has a cross section for neutron
capture $1.61 \times 10^4$ times that of H. Consequently, this isotope
is able to remove neutrons from the atmosphere rather efficiently, thereby
reducing the rate of their capture by protons. The density and the
composition also determine the mean free path of 2.22 MeV photons in the
atmosphere.

\begin{figure}
 \epsfig{file=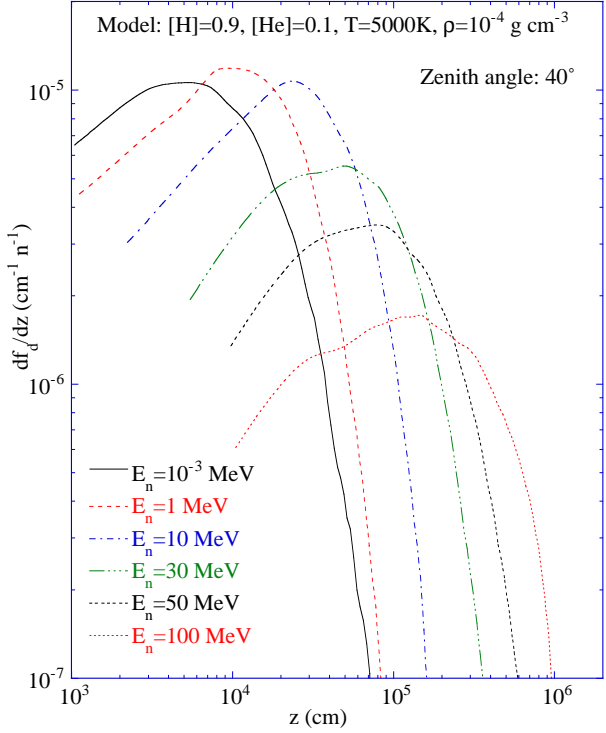,height=12.2cm,width=8.6cm}
 \caption{Distributions of the neutron capture sites in the secondary's
atmosphere  for several impinging neutron energies.}
 \label{fig:dfdz}
\end{figure}

Simulations of neutrons with an energy $E_n$ impinging on the atmosphere
at a zenith angle $\psi$ have also been performed. For each assumption
of E$_n$ and $\psi$, a depth distribution of neutron capture by H
($df_d(z;E_n,\psi)/dz$) and an energy and zenithal distribution of
neutrons leaving the atmosphere ($ dF(E,\theta;E_n,\psi) /(dE d\theta) $)
are computed. These
distributions are normalized to the number of impinging neutrons. Figure
\ref{fig:dfdz} shows the depth distribution of the sites of neutron capture
by protons for
neutrons impinging with an angle of 40 degrees. For neutron energies lower
than $\approx$ 0.01 MeV, the depth distribution of neutron capture sites
and the energy and zenithal distribution of neutrons leaving the
atmosphere are rather constant (with respect to energy) for a given initial
zenith
angle. In fact, at these energies, the neutrons are quickly thermalized in
the vicinity of the surface and all follow the same thermal diffusion.

%%%%%%%% Figure: Depth distribution of neutron capture %%%%%%%%%%%%%%
%%%%%%%%%%%%%%%%%%%%%%%%%%%%%%%%%%%%%%%%%%%%%%%%%%%%%%%%%%%

Depending on their energy and direction, some of the neutrons that are
ejected out of the atmosphere get decelerated gravitationally and fall
back onto the secondary's surface; some of them decay in the process. This
fraction ($F_r(E_n,\psi)$) of neutrons returning to the atmosphere depends
on the energy and zenith angle of the impinging neutrons. Indeed, the
larger the neutrons' energy, the deeper they can go, and the smaller the
zenith angle of the impinging neutrons the deeper they will also go.
Consequently, the probability for a neutron to leave the atmosphere
decreases with increasing energy and decreasing zenith angle. The fraction
$F_r(E_n,\psi)$ is calculated using Equation (19) of Hua \& Lingenfelter
(1987), which determines the time required for a neutron to return to the
atmosphere as a function of its energy and zenith angle. This time is
used to determine the fraction of neutrons decaying in flight before they
return to the atmosphere. The spectrum of the escaping and returning
neutrons has a cut-off value below the escape energy ($E_n^{(esc)} = {GMm_n
\over R} \approx 2 keV {{M/M_{\odot}} \over {R/R_{\odot}}}$) because even
if a neutron has a kinetic energy a little less than the escaping energy,
the time spent to return to the atmosphere can be larger than its lifetime.

The angular distribution of the returning neutrons is very close to an
isotropic distribution. Simulations have thus been performed to
derive the depth distribution of 2.22 MeV creation sites
($df_r(z)/dz$) due to returning neutrons. As previously stated,
this distribution is flat for low neutron energies. A fraction $F_{r,0}$ of
these returning neutrons can again
leave the secondary. However they all return to the atmosphere, since
their energies are smaller than the escaping energy, and they can then
again undergo one of the following: generate 2.22 MeV photons, decay, be
captured by $^3$He, or leave again the secondary atmosphere (with a
fraction given by $F_{r,0}$).

The total depth distribution for the radiative capture of neutrons is
obtained by adding to the ``direct" depth distribution (without
taking into account the returning neutrons) the returning-neutrons
component. This component is obtained by calculating the sum of the
geometric series of the fraction $F_{r,0}$. The total depth
distribution of 2.22 MeV creation site is then:

\begin{equation}
\frac{df(z;E_n,\psi)}{dz} =  \frac{df_d(z;E_n,\psi)}{dz} +
\frac{df_r(z)}{dz} \frac{F_r(E_n,\psi)}{(1-F_{r,0})}
\label{eq:p1}
\end{equation}

\begin{figure}
 \epsfig{file=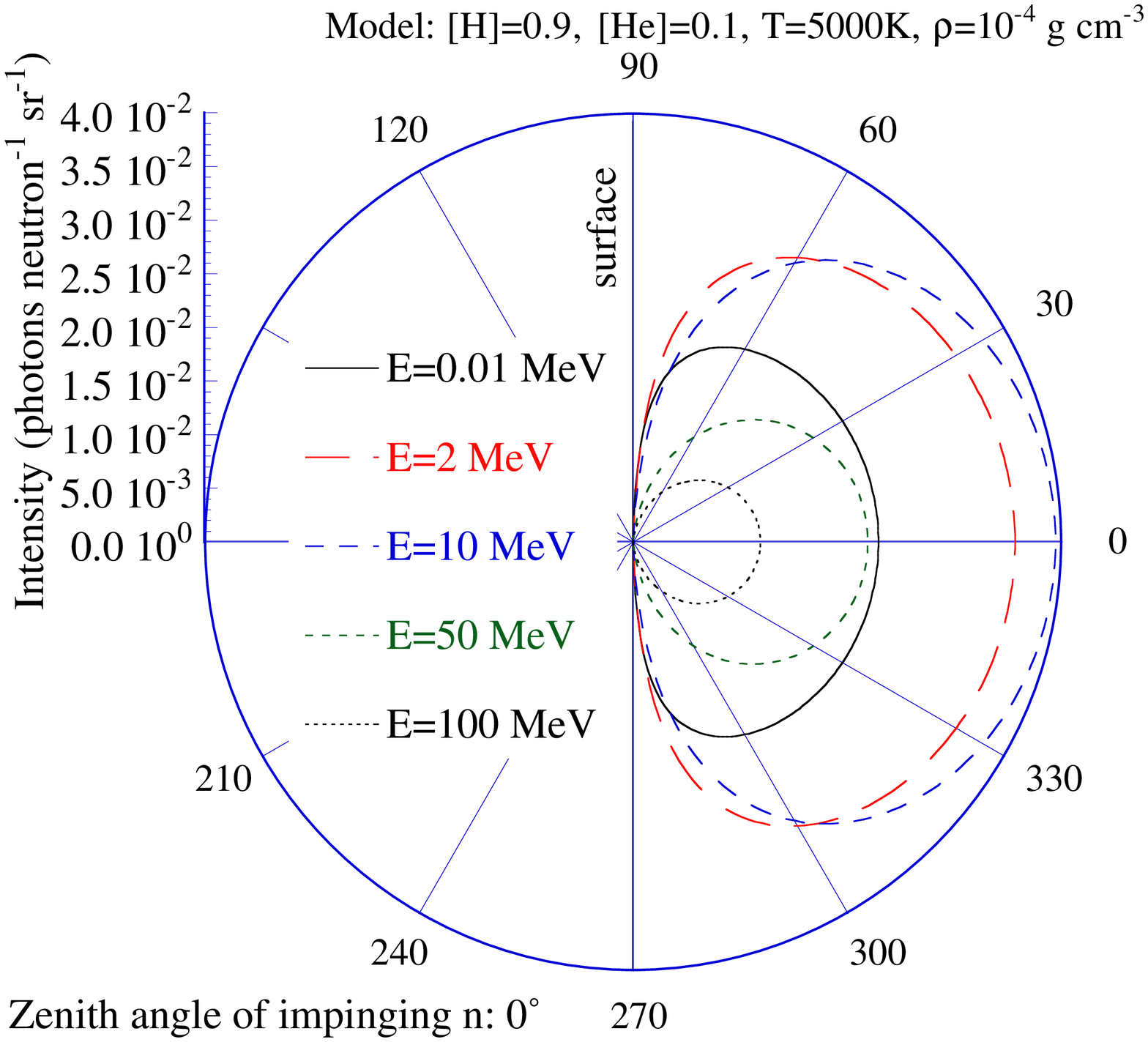,height=6.1cm,width=8.6cm}
 \caption{Angular distributions of the 2.22 MeV photons emitted from the
secondary atmosphere for several impinging neutron energies.}
 \label{fig:distang}
\end{figure}

\noindent The angular distribution of the escaping 2.22 MeV photons is computed
using the total depth distribution of 2.22 MeV photon creation sites. Let
$\xi$ and $\chi$ be the zenithal and azimutal direction angles of the 2.22
MeV photon. The angular distribution of escaping 2.22 MeV photons induced
by a neutron of energy $E_n$ impinging on the atmosphere with a zenith
angle of $\psi$ is:

\begin{equation}
\frac{d\epsilon_{2.2}(\xi,\chi;E_n,\psi)}{d\Omega} = \frac{1}{4\pi} \int_z
\frac{df(z;E_n,\psi)}{dz} \, e^{\frac{-z}{\lambda_{2.2}cos\xi}} \, dz
\label{eq:p2}
\end{equation}

\noindent with $\lambda_{2.2}$ being the mean free path of 2.22 MeV photons
in the
secondary's atmosphere. This angular distribution, expressed in photons
steradian$^{-1}$ neutron$^{-1}$, is independent of the azimuth. Figure
\ref{fig:distang} shows an example of 2.22 MeV photons angular distribution
for neutrons impinging with various $E_n$'s at a zenith angle of 0$^o$.

%%%%%%%% Figure: Angular distribution of 2.22 MeV photons %%%%%%%%%%%%%%
%%%%%%%%%%%%%%%%%%%%%%%%%%%%%%%%%%%%%%%%%%%%%%%%%%%%%%%%%%%

\begin{figure}
 \epsfig{file=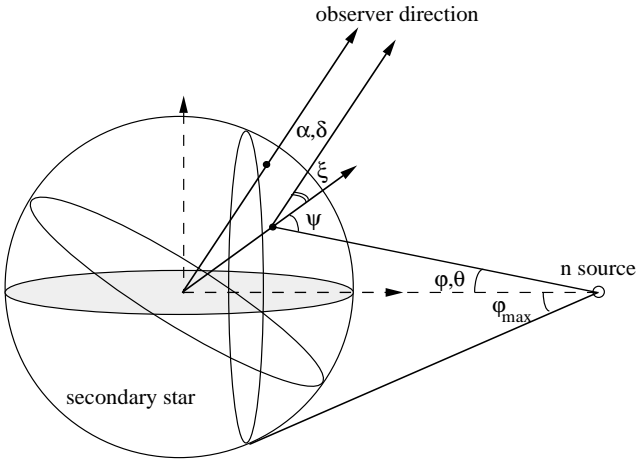,height=6.1cm,width=8.6cm}
 \caption{Schematic view of the irradiation of the secondary star by neutrons.
The binary system frame is represented by the dashed axes.}
 \label{fig:draw}
\end{figure}

The total 2.22 MeV intensity depends on the direction of observation with
respect to the binary system frame. Indeed, the 2.22 MeV photons are
emitted only from the surface of the secondary that is irradiated by
neutrons. Therefore, the observable gamma-ray flux comes only from the
fraction of the secondary area that is irradiated by neutrons and is
visible for the observer (see Fig. \ref{fig:draw}). So we can expect to
observe a periodic 2.2 MeV line flux due to the rotation of the binary
system. The total 2.22 MeV intensity $I_{2.2}(\alpha,\delta)$ (photons
s$^{-1}$
sr$^{-1}$) in a direction ($\alpha$,$\delta$) can be obtained by estimating
the integral:

\begin{equation}
I_{2.2}(\alpha,\delta) = \frac{N_n}{4\pi} \int_{D}
\int^{\infty}_{E_{c}} \frac{dF(E)}{dE}
\frac{d\epsilon_{2.2}(\xi,\chi;E,\psi)}{d\Omega} d\Omega dE
\label{eq:p3}
\end{equation}

\noindent where $dF(E)/dE$ is the energy distribution of neutrons
reaching the secondary, $N_n$ is the neutron rate (in n s$^{-1}$) emitted
by the accretion disk, $\varphi$ and $\theta$ the direction of the emitted
neutrons (see Fig. \ref{fig:draw}) and $D$ is the integration domain ($
\varphi , \theta \in D $) defined as
the intersection of the irradiated area (delimited in a cone of angle
$\varphi_{max}$) and the visible area in the direction $\alpha$ and $\delta$.
In this equation, $\xi$ is a function of $\varphi$, $\theta$, $\alpha$ and
$\delta$. The zenith angle $\psi$ depends only on $\varphi$. The integration
domain depends not only on the direction of emission but also on the
radius of the secondary and on its distance with respect to the neutron
source.

%%%%%%%% Figure: Geometry of the irradiation %%%%%%%%%%%%%%
%%%%%%%%%%%%%%%%%%%%%%%%%%%%%%%%%%%%%%%%%%%%%%%%%%%%%%%%%%%

%%%%%%%%%%%%%%%%%%%%%%%%%%%%%%%%%%%%%%%%%%%%%%%%%%%%%%%%%%%%%%%%%%%%%%%%%%
% 4. Results
%%%%%%%%%%%%%%%%%%%%%%%%%%%%%%%%%%%%%%%%%%%%%%%%%%%%%%%%%%%%%%%%%%%%%%%%%%

\section{\label{s4} Results}

We have considered two models for the composition and the geometrical
characteristics of the secondary star (see Table \ref{tab:model}). The
values shown in the table have been chosen so as to compare neutron capture
efficiency in
Helium-enriched
and non-enriched (``normal") atmospheres. Calculations of the 2.22 MeV
emissivity with specific, accurate secondary star parameters of
known binary systems will be presented in a separate publication (Jean et
al. 2001, in preparation).

\begin{table}
    \caption{Characteristics of the secondary star for the two models.}
    \begin{array}[b]{lcc}
\hline
\noalign{\smallskip}
                     & \; \; {\rm Model \; 1} \; \;  & \;  \; {\rm Model \; 2} \;  \\
\hline
\noalign{\smallskip}
{\rm Radius} \, (R_{\odot})   & \; 1 \;        & \;  3.6 \;      \\
{\rm Mass} \, (M_{\odot})     & \; 1 \;        & \; 10 \;         \\
{\rm Density} \, (g/cm^3)   & \; 10^{-4}\; & \;  10^{-5} \; \\
{\rm Temperature} \, (K)      & \; 5000 \;     & \;  10000  \;   \\
\mathrm{[H]}                  & \; 0.9  \;     & \;  0.1 \;       \\
\mathrm{[He]}                 & \; 0.1  \;     &  \; 0.9  \;     \\
\noalign{\smallskip}
\hline
    \end{array}
  \label{tab:model}
\end{table}

Calculations based on the method presented in the previous section have
been performed in the aim of deriving the expected flux in the 2.22 MeV
line from the binary system. The energy distribution of neutrons was
assumed to be a Maxwellian with kT = 10 MeV, truncated at 5 MeV ($E_c$ in
equation \ref{eq:p3}) because
the gravitationnal potential of the primary compact object retains the
neutrons of lower energy (see Sect. 2). Since we are dealing with close
binary systems, the separation is less than 40 R$_{\odot}$, which is the
mean free
path of 5 MeV neutrons in the vacuum. Therefore, the energy distribution
does not necessitate an additionnal cut to account for the loss of
neutrons by decay during their trip to the secondary star.

%%%%%%%%%%%%%%%%%%%%%%%%%%%%%%%%%%%%%%%%%%%%%%%%%%%%%%%%%%%%%%%%%%%%%%%%%%
\subsection{\label{s31} Fraction of escaping 2.22 MeV photons}

\begin{figure}
 \epsfig{file=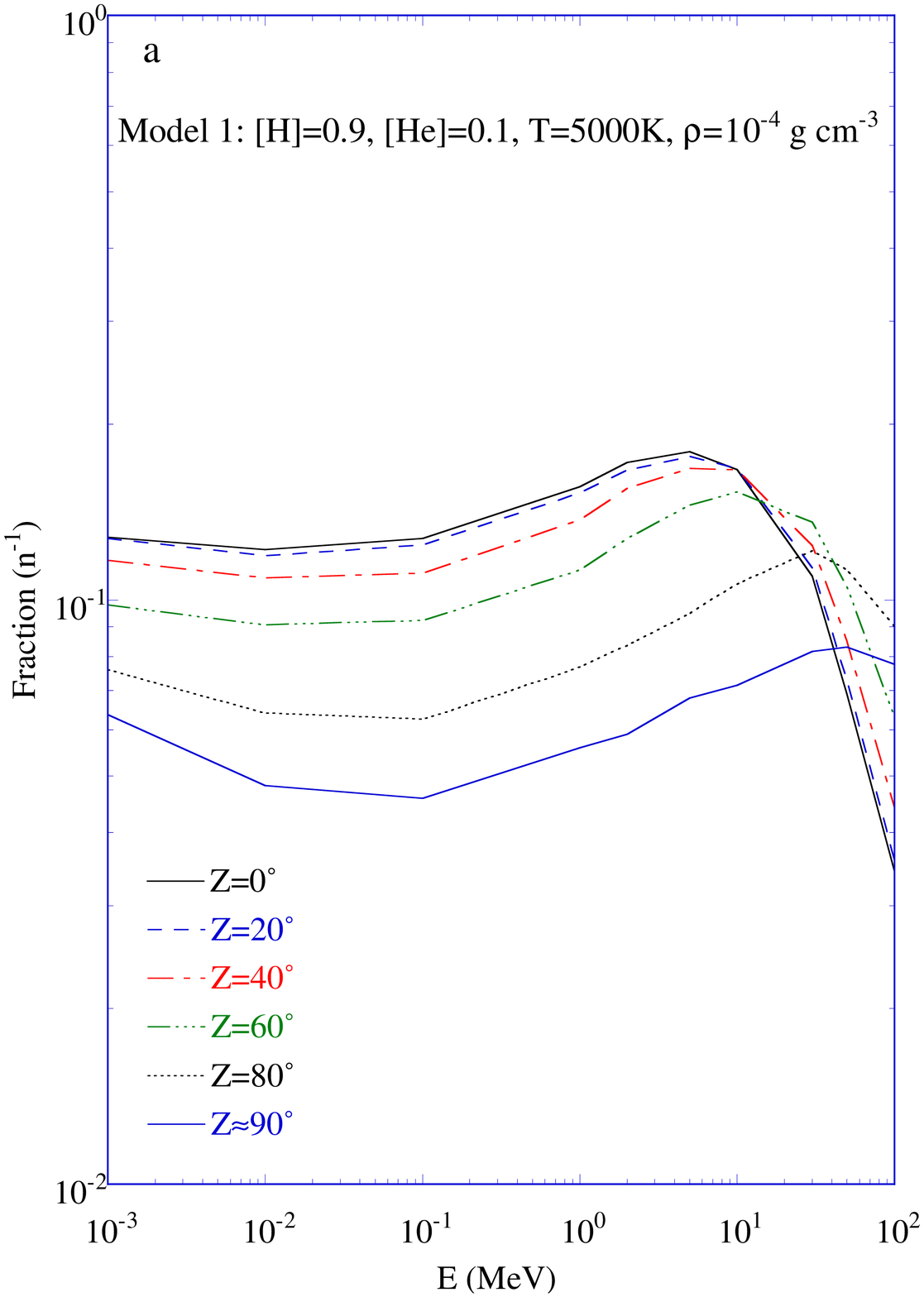,height=9.8cm,width=7cm}
 \epsfig{file=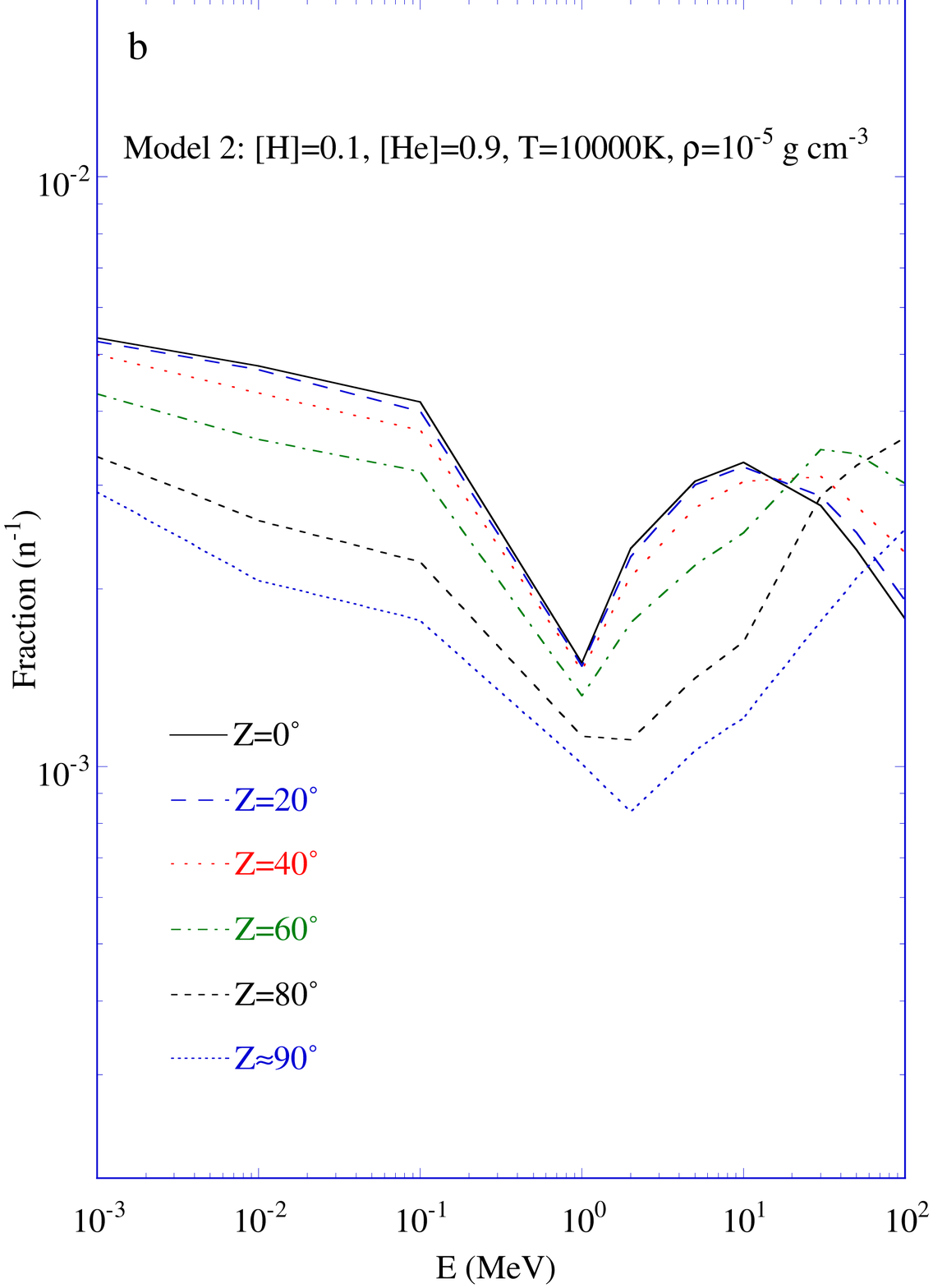,height=9.8cm,width=7cm}
 \caption{Fraction of escaping 2.2 MeV photons as a function of the
impinging neutron energies for several impinging zenith angle. The two
models are presented separately.}
 \label{fig:f22}
\end{figure}

The fraction of escaping 2.22 MeV radiation ($f_{2.2}$) is defined as the
ratio of 2.22 MeV photons that reach the secondary star surface (outward)
without scattering to the total number of incident neutrons. This fraction
is calculated by integrating the angular distribution of escaping 2.22 MeV
photons (see equation 4) over all the possible angles. The 2.22 MeV
photons produced by the returning neutrons are also taken into account. The
fraction $f_{2.2}$ depends on the zenith angles and energies of incident
neutrons. It gives an idea of the efficiency of neutrons in producing an
observable 2.22 MeV line. Figures \ref{fig:f22} show the escaping
2.22 MeV fraction for the two models of the secondary star. The difference
in the neutron capture rate between the two models is principally due to
the difference in the composition of the atmosphere. Indeed the abundance
of $^3$He in the second model is such that a large fraction of neutrons
are captured by this nucleus and thus lost. Consequently, the neutrons
capture rate decreases by a factor $\approx$ 17 between the two models. In
Model-2 the 2.22 MeV photons are emitted deeper in the atmosphere, and so
their shorter mean free path lowers the fraction of escaping photons
by a factor 1.5 compared to Model-1. Finally, the gravitational
escaping energy of neutrons in Model-2 is slighty lower than in Model-1,
increasing the fraction of escaping neutrons and further reducing the 2.22
MeV emissivity in Model-2. All these effects lead to a decrease of
$f_{2.2}$ by a factor $\approx$ 30 for Model-2 with respect to Model-1 at
low $E_n$.

%%%%%%%% Figures: 2.2 MeV escaping fraction %%%%%%%%%%%%%%
%%%%%%%%%%%%%%%%%%%%%%%%%%%%%%%%%%%%%%%%%%%%%%%%%%%%%%%%%%%

The fraction of escaping 2.22 MeV photons depends not only on the neutron
capture rate but also on the transmission of 2.22 MeV photons in the
atmosphere. In Model-1 (see Fig. \ref{fig:f22}a), where the neutron energy is
lower than the escaping energy, the neutrons thermalize quickly and stay
in the upper layer of the atmosphere. If they leave the atmosphere,
they cannot escape the secondary but a fraction of them decay before
returning in the atmosphere. The fraction of escaping neutrons
increases with the zenith angle. Indeed, high-zenith-angle incident
neutrons can more easily escape since they scatter close to the surface.
This is not the case for
low-zenith-angle incident neutrons that go deeper in the atmosphere: even
if they scatter toward the surface, their energies are
rarely above the escape value. This explains why the $f_{2.2}$ fraction
decreases with increasing zenith angle below 5 MeV. Above 5
MeV, neutrons penetrate and are captured deeper in the atmosphere. The
fraction of escaping neutrons is reduced, as well as the transmission of
2.22 MeV photons. However, for large zenith angles, the neutron captures
happen closer to the surface in zones allowing the 2.22 MeV photons to
escape more easily from the star. 

The variation of the 2.22 MeV photon
escaping fraction as a function of the angle and the energy of impinging
neutrons in Model-2 is similar to that in Model-1. However, a resonance in the
elastic scattering of neutrons with $^4$He at 1.1 MeV is at the origin of
the drop around 1 MeV (see Fig. \ref{fig:f22}b). At this resonance, the
angular distribution of $^4$He-scattered neutrons is peaked for backward
directions \footnote{see the ENDF/B-VI evaluated neutron data; http://t2.lanl.gov/cgi-bin/nuclides/endind}. Consequently, around 1 MeV, incident neutrons are more likely to backscatter and to be ejected out of the atmosphere with kinetic energies too high to allow them to return. This effect reduces the number of MeV neutrons that can thermalize and produce 2.2 MeV photons, hence the trough seen in Fig. \ref{fig:f22}b.

%%%%%%%%%%%%%%%%%%%%%%%%%%%%%%%%%%%%%%%%%%%%%%%%%%%%%%%%%%%%%%%%%%%%%%%%%%

\subsection{\label{s32} Flux versus model and irradiation geometry}

The distance between the accretion disk (the source of neutrons) and the
center of the secondary star, and the angles $\alpha$ \& $\delta$, which
specify the direction of Earth (see Fig. \ref{fig:draw}), are the basic
parameters of
this calculation. Since the binary system is rotating, $\alpha$ varies with
time
and the observed flux becomes periodic. Intensities have been calculated as
a function of the
direction of emission for different values of the distance between the
primary and secondary stars. Assuming parameters of the binary system
(separation and
masses), it has been possible to derive light curves of the 2.22 MeV line
emission, depending on $\delta$. However, the range of interesting
separations is limited since the secondary radius needs to fill its Roche
lobe in order to allow the accretion of its matter by the compact object.
Using a study by Paczynski (1971) that calculated the critical radius of a
star in a binary system that allows the outward flow of its matter, we
estimated the upper limit of the separation to be between 3.5 - 6
R$_{\odot}$ for a secondary star of 1 M$_{\odot}$ (Model-1) and 7.5 - 11
R$_{\odot}$ for a secondary star of 10 M$_{\odot}$ (Model-2), providing
compact object masses ranging from 3 to 20 M$_{\odot}$. With this type of
close binary systems, the period of rotation is less than a day.

\begin{figure}
 \epsfig{file=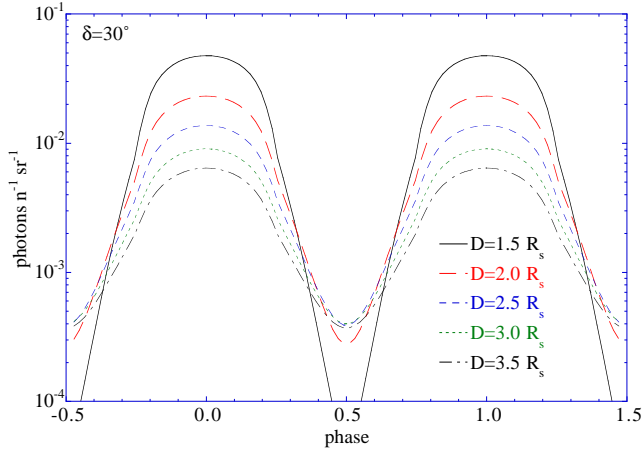,height=6.1cm,width=8.6cm}
 \caption{Example of phasogram of the 2.2 MeV emission for the Model-1 with
a compact object mass of 3 M$_{\odot}$. $\delta$ is the angle between the
binary system plan and the direction of the observer. D is the distance
between the accretion disk and the secondary star in solar radius unit. The
period of the binary system change from 0.1 to 0.4 days.}
 \label{fig:phaso}
\end{figure}

Figure \ref{fig:phaso} shows phasograms of the 2.22 MeV intensity as a
function of
the separation for Model-1. As expected, the intensity is maximum when the
neutron-irradiated area of the secondary star atmosphere faces the
observer. The maximum value of the intensity decreases with increasing
separation values because the neutrons flux impinging on the atmosphere
decreases with increasing distance between the accretion disk and the
secondary. In Fig. \ref{fig:phaso}, the intensity is equal to zero at a phase
value of 0.5 only for a separation of 1.5 R$_{\odot}$, because for this
value of
$\delta$ the irradiated area is hidden by the star. For larger
separations, the irradiated area becomes larger and a fraction of this
area is always visible by the observer.

%%%%%%%% Figures: Phasogram %%%%%%%%%%%%%%
%%%%%%%%%%%%%%%%%%%%%%%%%%%%%%%%%%%%%%%%%%%%%%%%%%%%%%%%%%%

\begin{figure}
 \epsfig{file=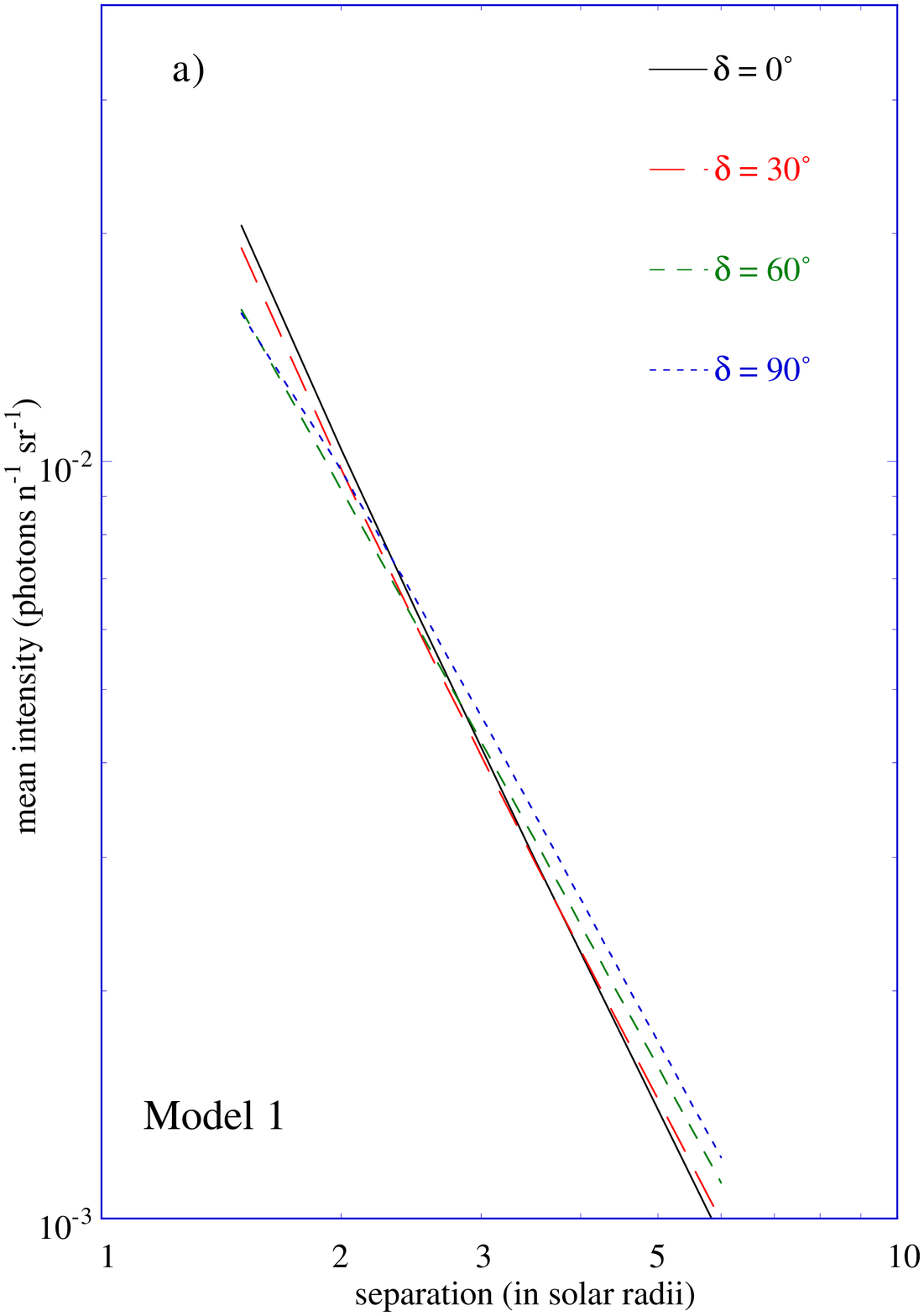,height=9.8cm,width=7cm}
 \epsfig{file=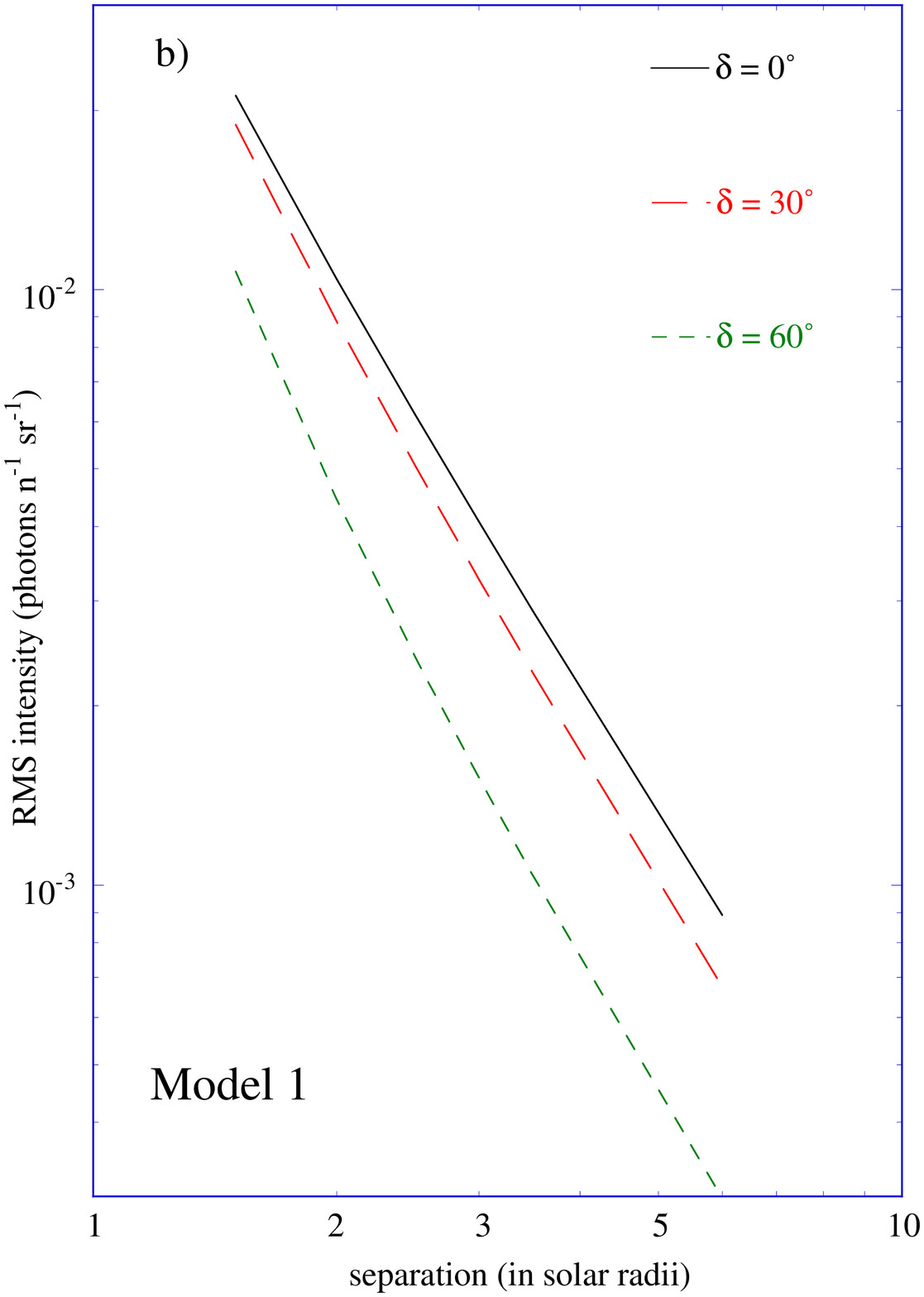,height=9.8cm,width=7cm}
 \caption{Normalized mean (a) and RMS (b) intensities for Model-1 as a function
of the separation of the binary. Several values of $\delta$ are shown.}
 \label{fig:intens1}
\end{figure}

\begin{figure}
 \epsfig{file=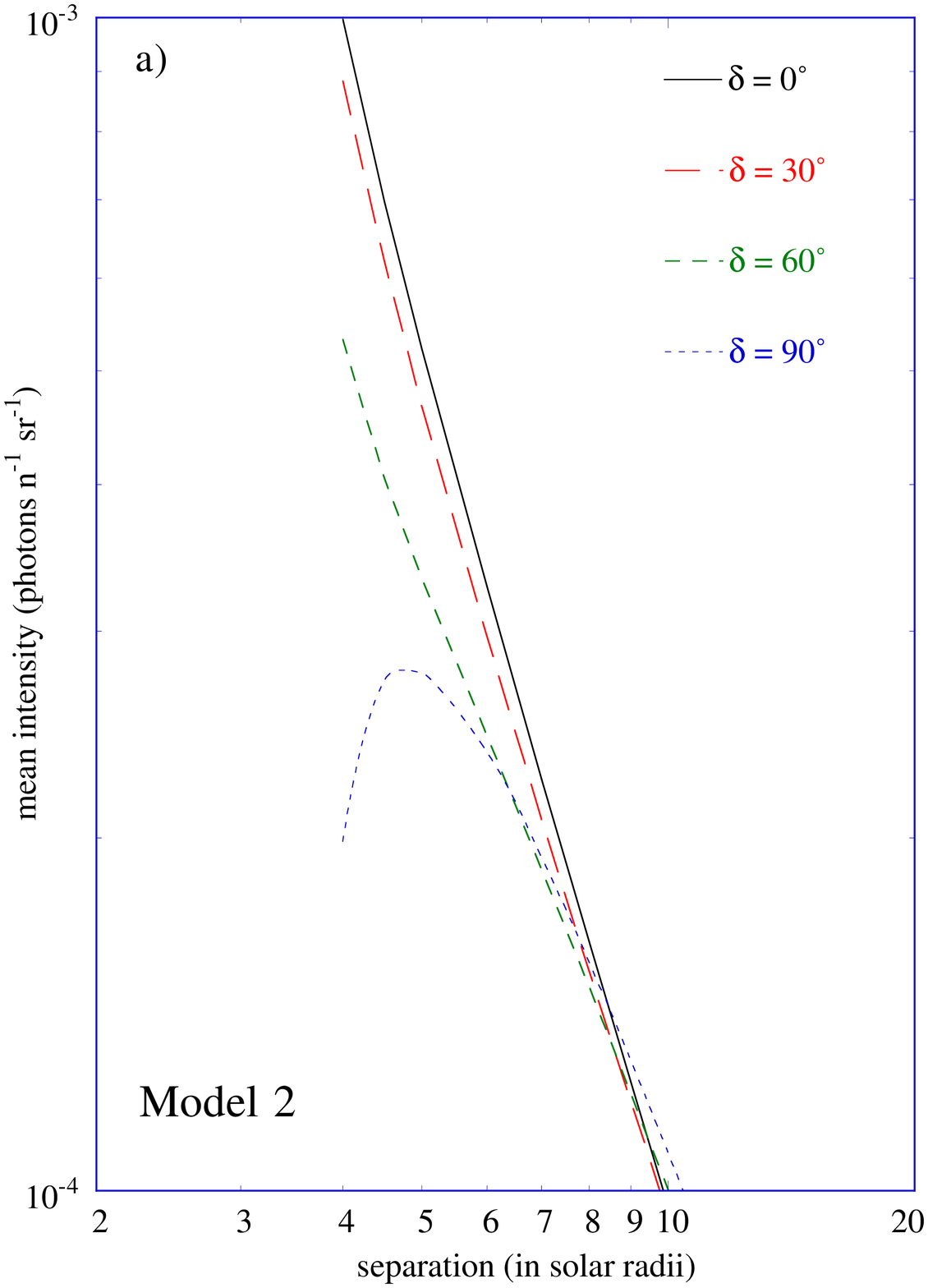,height=9.8cm,width=7cm}
 \epsfig{file=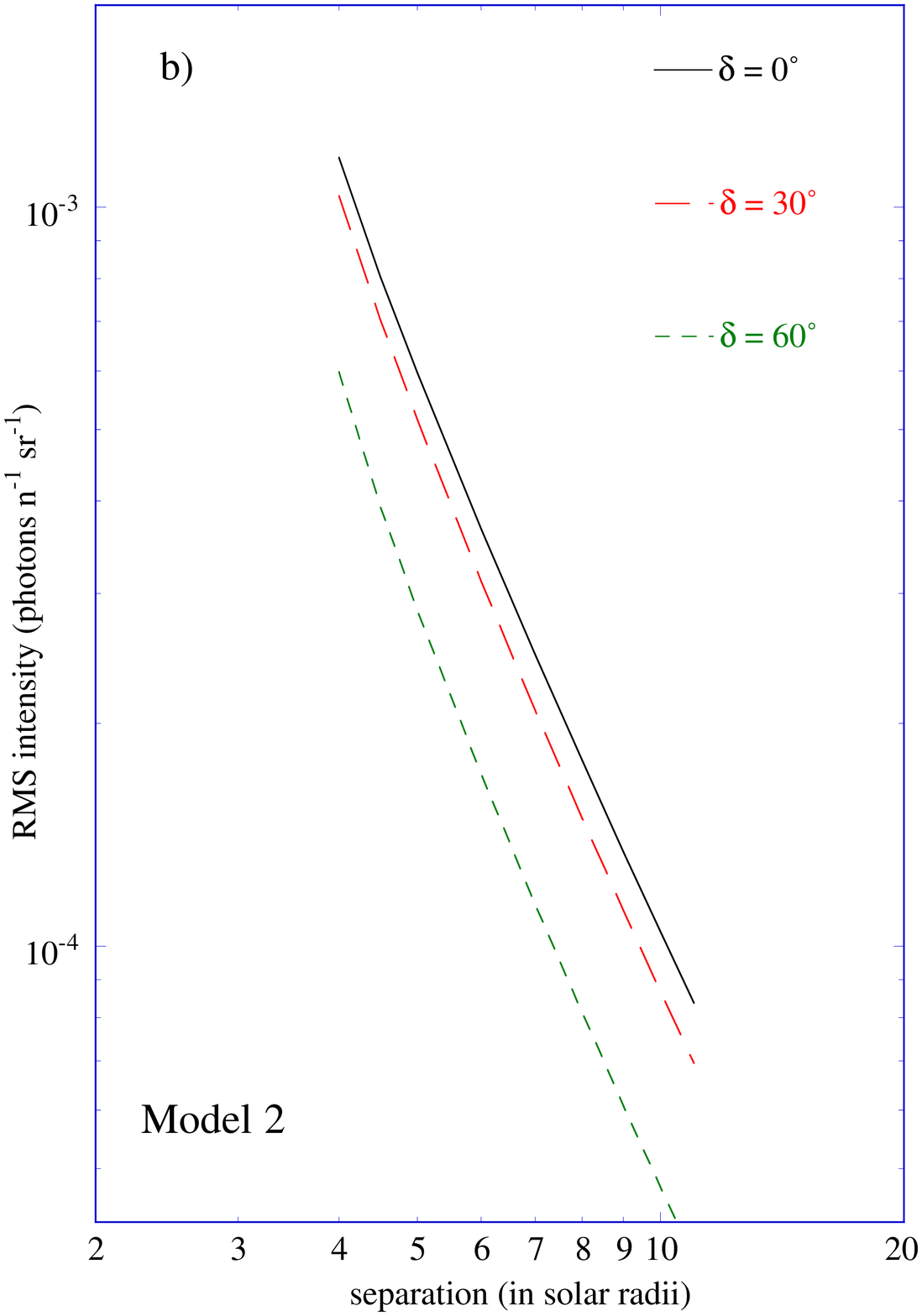,height=9.8cm,width=7cm}
 \caption{Normalized mean (a) and RMS (b) intensities for Model-2 as a function
of the separation of the binary. Several values of $\delta$ are shown.}
 \label{fig:intens2}
\end{figure}

The mean and the root-mean-square (RMS) fluxes are commonly used for the
analysis of periodic emissions. They have been estimated for our 2.22 MeV
intensity as a function of the separation and the direction of the
observer. Both the mean and the RMS intensities decrease with increasing
separation values as a power law with slope around -2 (between -1.8 and
-2.3). Figures \ref{fig:intens1} and  \ref{fig:intens2} show the variation
of the mean and RMS intensities as a function of the binary separation for
the two models.

The mean intensity does not vary significantly with the angle of the
Observer's direction with respect to the binary system plane. The
RMS intensity decreases, however, showing a reduction of the modulation of the
visible photon flux with increasing $\delta$. The modulation
obviously disappears for a direction of observation of the binary system
perpendicular to the plane of the latter ($\delta$=90$^o$).

%%%%%%%% Figures: 2.2 MeV mean and rms intensities Model 1 %%%%%%%%%%%%%%
%%%%%%%%%%%%%%%%%%%%%%%%%%%%%%%%%%%%%%%%%%%%%%%%%%%%%%%%%%%

%%%%%%%% Figures: 2.2 MeV mean and rms intensities Model 2 %%%%%%%%%%%%%%
%%%%%%%%%%%%%%%%%%%%%%%%%%%%%%%%%%%%%%%%%%%%%%%%%%%%%%%%%%%

The flux at 2.22 MeV can be derived using the formula:

\begin{equation}
F_{2.2} = 10^{-5} \, \frac{R_n}{10^{40} \, n \, s^{-1}} \,
\frac{I_{2.2}}{10^{-2} \, n^{-1} \, sr^{-1}}  \, d^{-2}_{kpc} \, s^{-1} \,
cm^{-2}
\label{eq:p4}
\end{equation}

\noindent where $R_n$ is the rate of neutrons emitted by the accretion disk,
$I_{2.2}$ is the normalized intensity and $d_{kpc}$ the distance of the
binary system to the observer in kpc. For Model-1, the flux at 1 kpc is
found to
be in the range 10$^{-7}$ -- 10$^{-5}$ photons s$^{-1}$ cm$^{-2}$
depending on the accretion disk model. For Model-2, it is found to be
in the range 10$^{-8}$ to 10$^{-6}$ photons s$^{-1}$ cm$^{-2}$
depending on the accretion disk model. Combining the results of $R_n$ from
Sect. 2 (Tables 1a,b to 5a,b) with the calculations in this section, we
obtain the final fluxes $F_{2.2}$, which we present succinctly in Tables
7 and 8. The separations have been chosen to be 2 R$_{\odot}$ and 8 R$_{\odot}$
for Model-1 and Model-2 respectively, since for these values the mean
fluxes do not change significantly with the binary system inclination (see
Figs. \ref{fig:intens1} and \ref{fig:intens2}).

\begin{table}
    \caption{Mean 2.22 MeV flux vs. Accretion disk models (at 1 kpc for a
separation of 2 R$_{\odot}$ and an initial 90\% H and 10\% He Composition) }
    \begin{array}[b]{lcccc}
\noalign{\smallskip}
{\rm Accretion \, model} & \dot M \, (M_{\odot}/yr) & \alpha=0.3 &
\alpha=0.1 \\
\hline
 \noalign{\smallskip}
        & 10^{-10}   &  1.1\times 10^{-8}  &  5.5\times 10^{-8}  \\
{\rm ADAF}    & 10^{-9}     &  4.3\times 10^{-7}  &  1.4\times 10^{-6}  \\
        & 10^{-8}     &  9.1\times 10^{-6}  &  1.5\times 10^{-5}  \\
        &               &                       &                       \\
        & 10^{-10}    &  1.6\times 10^{-8}  &  1.3\times 10^{-7}  \\
{\rm ADIOS}   & 10^{-9}     &  1.5\times 10^{-6}  &  7.8\times 10^{-6}  \\
        & 10^{-8}     &  8.3\times 10^{-5}  &  1.1\times 10^{-4}  \\
        &               &                       &                       \\
        & 10^{-10}    &  3.5\times 10^{-22} &  4.8\times 10^{-12} \\
{\rm SLE}     & 10^{-9}     &  4.1\times 10^{-9}  &  2.8\times 10^{-7}  \\
        & 10^{-8}     &  4.8\times 10^{-6}  &  1.8\times 10^{-5}  \\
        &               &                       &                       \\
	& 		 & kT_e=0.5 MeV & kT_e=0.1 MeV \\
        &               &                       &                       \\
        & 10^{-10}    &  3.1\times 10^{-7}  &  4.7\times 10^{-8}  \\
{\rm UIT-30 MeV} & 10^{-9}  &  2.3\times 10^{-6}  &  4.2\times 10^{-7}  \\
        & 10^{-8}    &  1.6\times 10^{-5}  &  4.9\times 10^{-6}  \\
        &               &                       &                       \\
        & 10^{-10}    &  1.9\times 10^{-7}  &  2.0\times 10^{-8}  \\
{\rm UIT-10 MeV} & 10^{-9}  &  1.6\times 10^{-6}  &  2.0\times 10^{-7} \\
        & 10^{-8}   &  1.1\times 10^{-5} &  1.7\times 10^{-6}  \\
 \noalign{\smallskip}
 \hline
    \end{array}
  \label{tab:6a}
\end{table}

\begin{table}
    \caption{Mean 2.22 MeV flux vs. Accretion disk models (at 1 kpc for a
separation of 8 R$_{\odot}$ and an initial 10\% H and 90\% He Composition) }
    \begin{array}[b]{lcccc}
\noalign{\smallskip}
{\rm Accretion \, model} & \dot M \, (M_{\odot}/yr) & \alpha=0.3 &
\alpha=0.1 \\
\hline
 \noalign{\smallskip}
        & 10^{-10}   &  1.0\times 10^{-9}  &  4.0\times 10^{-9}  \\
{\rm ADAF}    & 10^{-9}     & 3.9\times 10^{-8} &  9.2\times 10^{-8}  \\
        & 10^{-8}     &  6.0\times 10^{-7} & 1.1\times 10^{-6}  \\
        &               &                       &                       \\
        & 10^{-10}    &  2.1\times 10^{-9} & 1.2\times 10^{-8}  \\
{\rm ADIOS}   & 10^{-9}     &  1.3\times 10^{-7} & 5.2\times 10^{-7}  \\
        & 10^{-8}     & 5.4\times 10^{-6} & 1.5\times 10^{-5} \\
        &               &                       &                       \\
        & 10^{-10}    & 9.7\times 10^{-23} & 1.3\times 10^{-12} \\
{\rm SLE}     & 10^{-9}     & 7.5\times 10^{-10} & 2.7\times 10^{-8} \\
        & 10^{-8}     & 4.9\times 10^{-7} & 2.0\times 10^{-6} \\
        &               &                       &                       \\
	& 		 & kT_e=0.5 MeV & kT_e=0.1 MeV \\
        &               &                       &                       \\
        & 10^{-10}    & 1.1\times 10^{-8} & 1.5\times 10^{-9} \\
{\rm UIT-30 MeV} & 10^{-9}  & 9.0\times 10^{-8} & 1.5\times 10^{-8} \\
        & 10^{-8}    & 6.6\times 10^{-7} & 4.6\times 10^{-7} \\
        &               &                       &                       \\
        & 10^{-10}    & 6.0\times 10^{-9}  & 6.1\times 10^{-10} \\
{\rm UIT-10 MeV} & 10^{-9}  & 5.4\times 10^{-8} & 7.2\times 10^{-9}\\
        & 10^{-8}   & 4.1\times 10^{-7} & 1.2\times 10^{-7} \\
 \noalign{\smallskip}
 \hline
    \end{array}
  \label{tab:6b}
\end{table}

The rotation of the binary system leads to a shift in the centroid of the
2.22 MeV
line. This Doppler shift changes with the phase and depends on the
observer's direction ($\delta$). Using classical relations
in binary systems (e.g.~Franck, King \& Raine, 1992) we derive a relation
(Equation \ref{eq:p5}) for the spectral shift of the 2.22 MeV line
which depends on the masses of the compact object and the secondary
($M_{c}$ and $M_s$,
respectively, in solar masses), the separation between the two objects
(d$_s$ in solar radii),
the phase ($\phi$), and the angle between the observer's direction and the
binary system ($\delta$).

\begin{equation}
\Delta E = 3.2 \, \frac{M_c}{\sqrt{d_s(M_c+M_s)}} \, cos\delta \, sin\phi
\; keV .
\label{eq:p5}
\end{equation}

\begin{figure}
 \epsfig{file=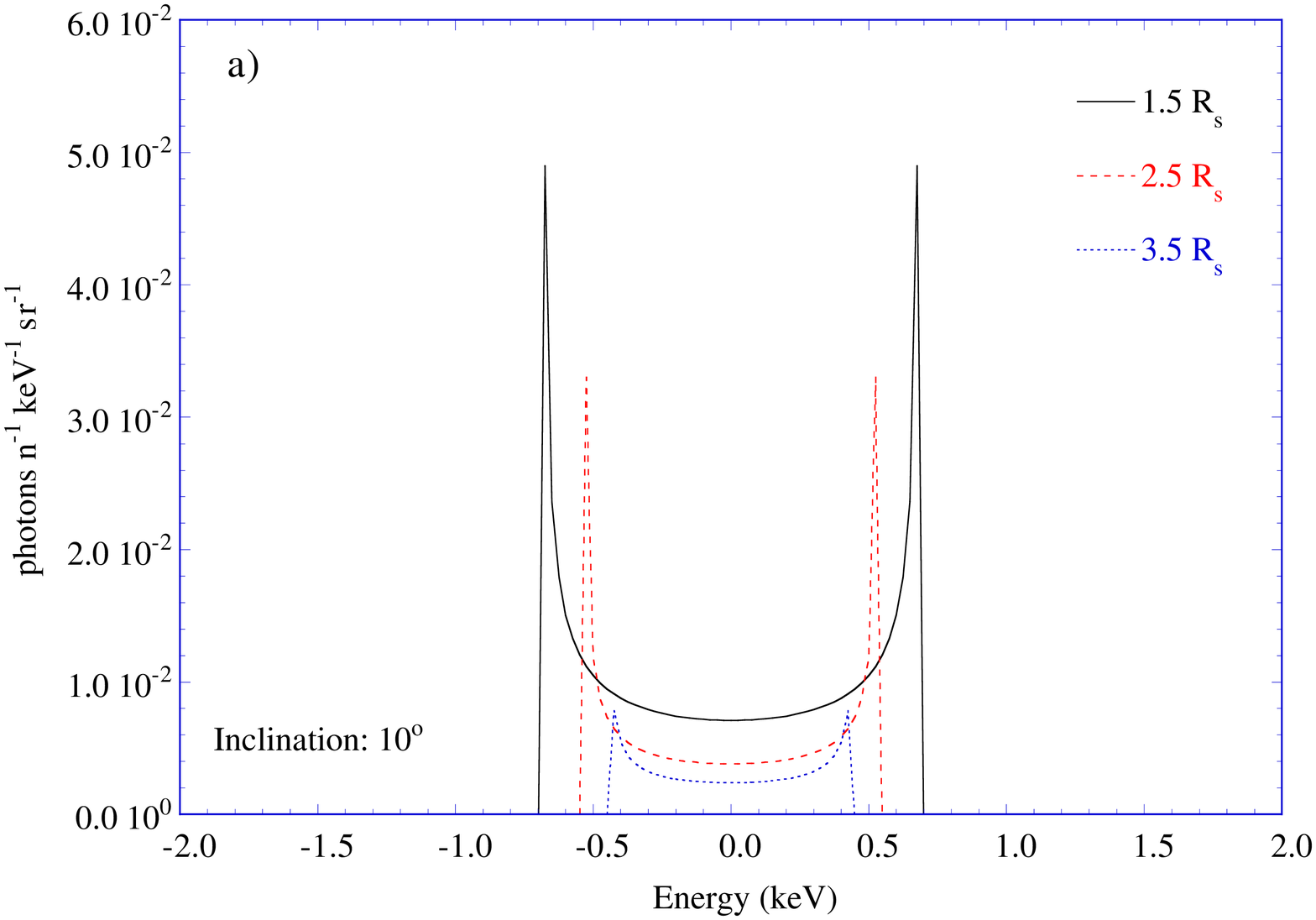,height=6.1cm,width=8.6cm}
 \epsfig{file=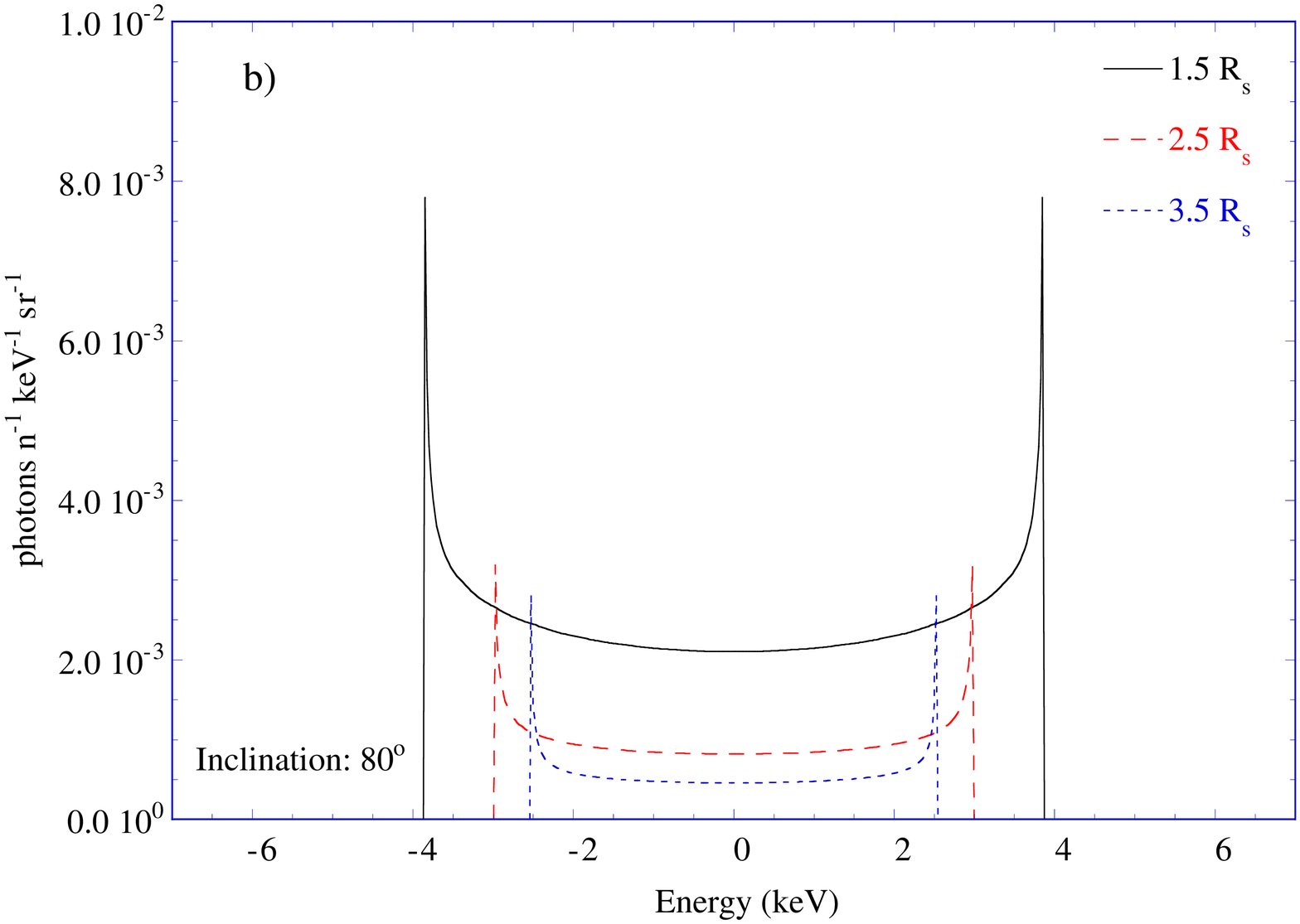,height=6.1cm,width=8.6cm}
 \caption{Shape of the neutron capture line for Model-1. The
inclination of the binary system is 10$^o$ (a) and 80$^o$ (b). Several values of
the binary separation are shown.}
 \label{fig:line-p}
\end{figure}

\begin{figure}
 \epsfig{file=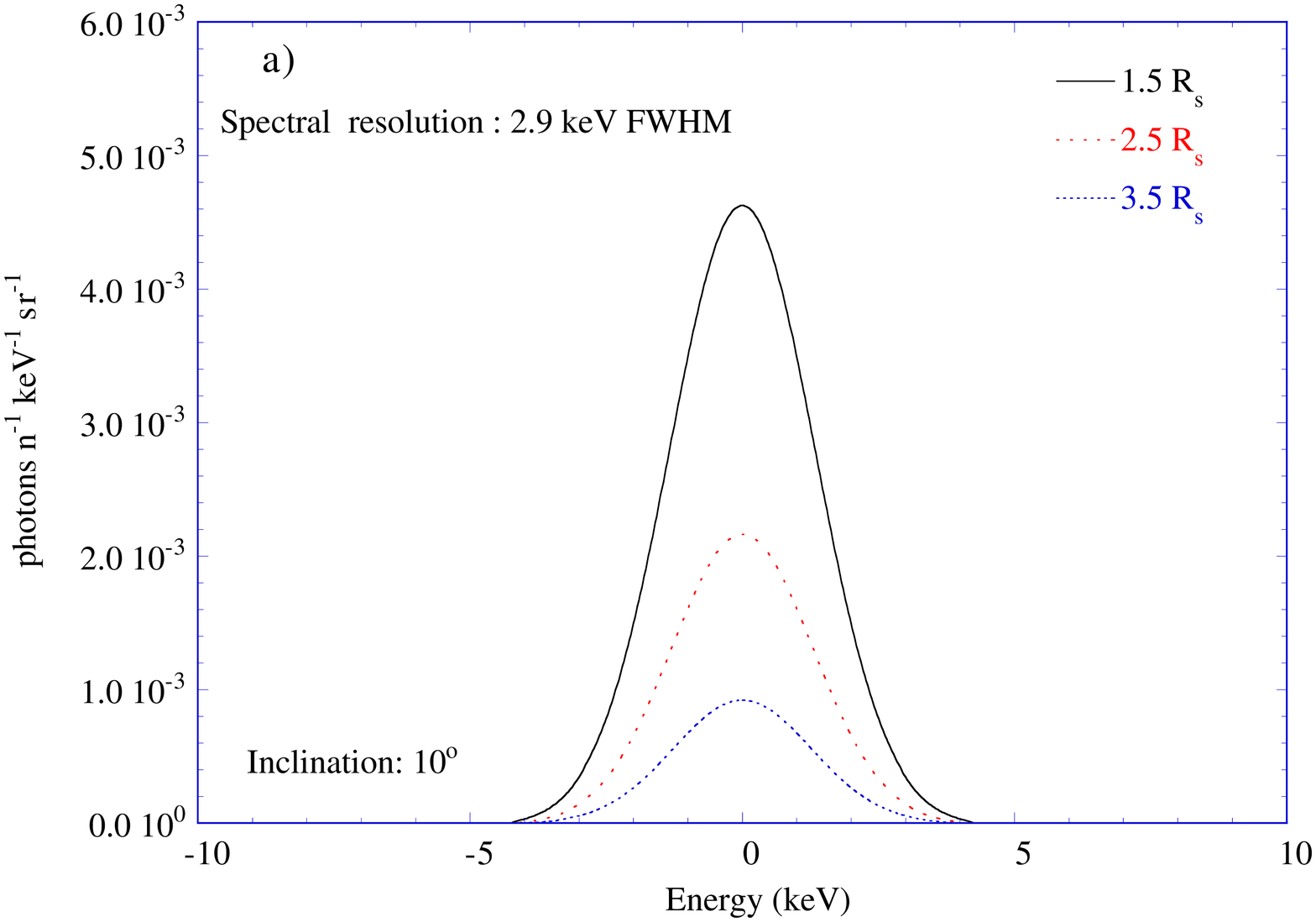,height=6.1cm,width=8.6cm}
 \epsfig{file=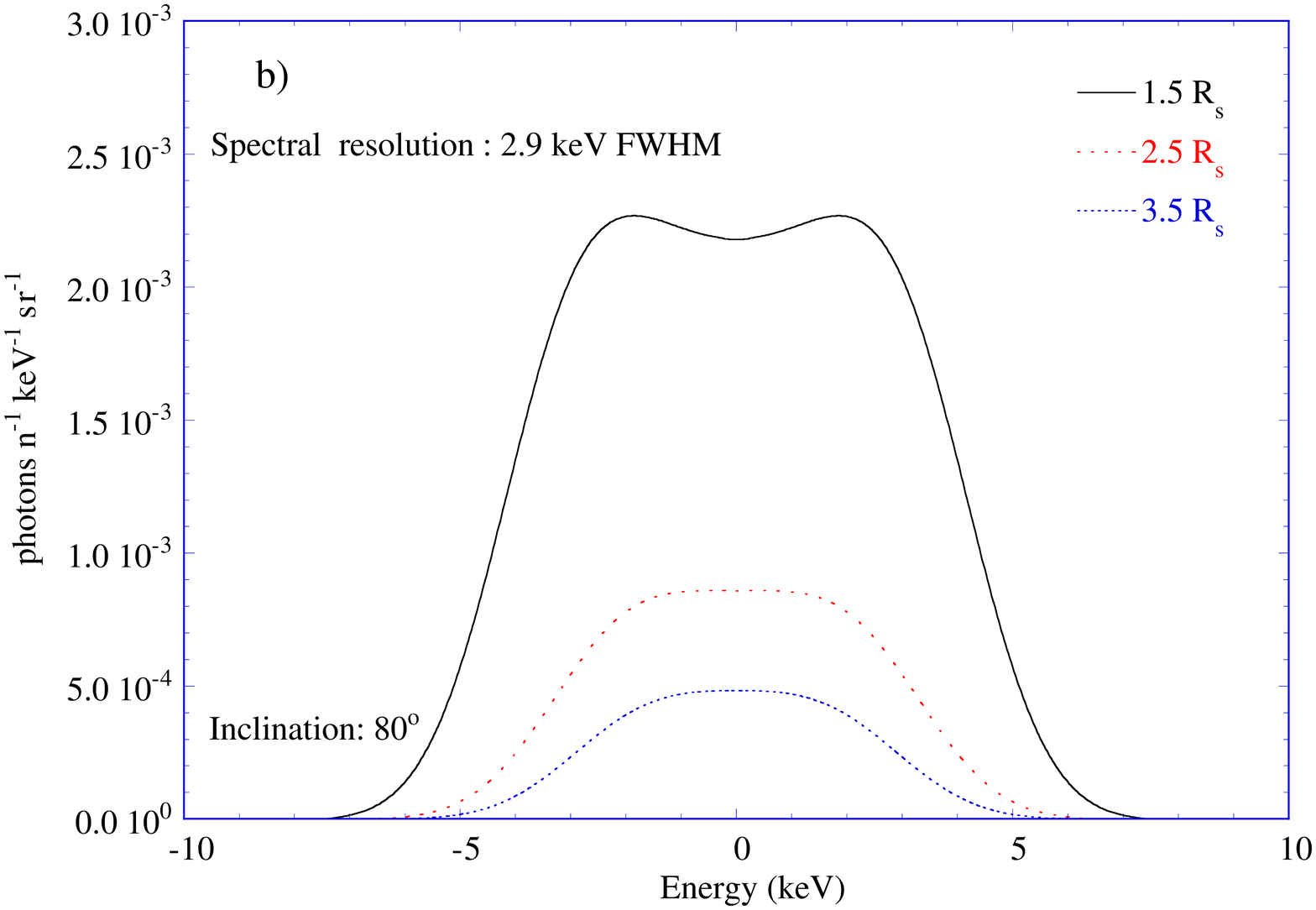,height=6.1cm,width=8.6cm}
 \caption{Estimation of the neutron capture line shape for Model-1 as
it could be measured by the spectrometer SPI of INTEGRAL. The inclination of
the binary system is 10$^o$ (a) and 80$^o$ (b). Several values of the binary separation are shown.}
 \label{fig:line-m}
\end{figure}

\noindent The estimated spectral shift is mesurable with SPI, the
spectrometer onboard the INTEGRAL satellite. Using Equation \ref{eq:p5} and
the intensity variation as a function of the phase (as presented in Figure
\ref{fig:phaso}) we have obtained the 2.22 MeV line profiles for Model-1
(Figures \ref{fig:line-p}) with a compact companion of 3 M$_{\odot}$ and
for two values of the binary system inclination\footnote{The inclination
$i$ is the angle between the binary system rotation axis with respect to
the observer direction, i.e. $i=90^o-\delta$.}. The line is broadened and
double-peaked due to the rotation of the secondary star. Figures
\ref{fig:line-m} show the shape of the line as it could be measured by SPI
(assuming a spectral resolution of 2.9  keV at 2.22 MeV). For an
inclination of 80$^o$ the double-peak shape is clearly visible, whereas
for a 10$^o$ inclination that is not the case.

%%%%%%%% Figures: Line shape - theoritical profile %%%%%%%%%%%%%%
%%%%%%%%%%%%%%%%%%%%%%%%%%%%%%%%%%%%%%%%%%%%%%%%%%%%%%%%%%%

%%%%%%%% Figures: Line shape - predicted profile %%%%%%%%%%%%%%
%%%%%%%%%%%%%%%%%%%%%%%%%%%%%%%%%%%%%%%%%%%%%%%%%%%%%%%%%%%

%%%%%%%%%%%%%%%%%%%%%%%%%%%%%%%%%%%%%%%%%%%%%%%%%%%%%%%%%%%%%%%%%%%%%%%%%%
% 5. Conclusions
%%%%%%%%%%%%%%%%%%%%%%%%%%%%%%%%%%%%%%%%%%%%%%%%%%%%%%%%%%%%%%%%%%%%%%%%%%

\section{\label{s5} Conclusions}

We have presented a theoretical investigation of the 2.22 MeV line emitted
from X-ray binary systems neutrons are captured in the secondary's
atmosphere. These neutrons are produced in the accretion
disk by nuclear spallation of Helium nuclei. The rate of production of neutrons
depends on various parameters set by the accretion disk model, i.e.~the
temperature,
density, and composition of the accreted material. Calculations of the line
intensity normalized to the neutron production rate have been performed for
two simple secondary models and for several X-ray binary geometries.

We showed that, due to the rotation of the
secondary, the intensity of the 2.22 MeV radiation is periodic, and
the line centroid is
shifted by the Doppler effect. The mean intensity does not vary
significantly with the direction of observation with respect to the binary
system frame.

According to our model, the spectral (Figures \ref{fig:line-p} and
\ref{fig:line-m}) and temporal
(Figures \ref{fig:phaso}, \ref{fig:intens1} and \ref{fig:intens2}) analyses
of the 2.22 MeV line flux would provide valuable insights into the
characteristics of the binary system (separation, inclination with respect
to the observer, masses, accretion rate and disk structure, etc.). If one
further knows the X-ray luminosity, the geometry of the X-ray binary
system, and its distance from Earth, then a measure of a 2.22 MeV line flux
from it can set constraints on the neutron production rate and consequently
on the accretion disk models (ADAF, ADIOS, SLE, etc.).

At 2.22 MeV the narrow $\gamma$-ray line sensitivity of the SPI spectrometer
 of the INTEGRAL mission is expected to be 7-10 $\times 10^{-6}$ photons
s$^{-1}$ cm$^{-2}$ for an observation time of 10$^6$ seconds (Jean et al.,
1999). The 2.22 MeV fluxes estimated for the simple secondary-star models
and for the various accretion disk models used here are found to be
measurable by the spectrometer SPI in some cases: mostly if the accretion
rate is large enough ($\dot M \approx$ 10$^{-8} M_{\odot}$/yr) and the
viscosity is significant ($\alpha \gsim 0.2$).
In case of a detection, the SPI spectral resolution
($\approx$ 3 keV at 2.22 MeV) would allow the measurement of the broadening
of the line, which is due to the rotation of the secondary, if the
separation  and the inclination of the X-ray binary are not too large.

In a future work, the 2.22 MeV line flux and shape will be estimated for
known X-ray binary systems (e.g. A0620-00, etc.) using all accurate,
available information on their characteristics (distance, separation,
composition, etc.). The calculated fluxes will be compared with upper-limit
fluxes obtained with
COMPTEL (Van Dijk, 1996), and correlation with other gamma-ray lines will be
investigated. Indeed, the neutrons that irradiate the secondary star can
also produce Be and Li isotopes in its atmosphere. This process was proposed
by Guessoum \& Kazanas (1999) to explain the overabundance of Li in some
soft X-ray transients. Both $^7$Li and $^7$Be emit gamma-ray lines (at
0.478 and 0.429 MeV), and $^7$Be decays into $^7$Li$^*$ (with a half-life
of 56 days), the latter then produces 0.478 MeV radiation upon de-exciting;
such a delayed gamma-ray emission, although weak, would be a clear
signature of the process. We can thus expect a certain correlation of
the 2.22 MeV emission with these gamma-ray lines, providing supplementary
insights into the physics of the binary system.

%%%%%%%%%%%%%%%%%%%%%%%%%%%%%%%%%%%%%%%%%%%%%%%%%%%%%%%%%%%%%%%%%%%%%%%%%%
\begin{acknowledgements}
We thank D. Kazanas for his useful comments. N. Guessoum would like to acknowledge a short visiting professor invitation
by Universit\'e Paul Sabatier and the American University of Sharjah for a summer research grant; he also wishes to thank Prs. P. von Ballmoos and
G. Vedrenne for the facilities provided at the Centre d'Etude Spatiale des
Rayonnements (Toulouse, France) where much of this work was conducted.
\end{acknowledgements}

%%%%%%%%%%%%%%%%%%%%%%%%%%%%%%%%%%%%%%%%%%%%%%%%%%%%%%%%%%%%%%%%%%%%%%%%%%

\end{document}